\documentclass[10pt,journal,letterpaper,compsoc]{IEEEtran}
\makeatletter
\def\ps@headings{%
\def\@oddhead{\mbox{}\scriptsize\rightmark \hfil \thepage}%
\def\@evenhead{\scriptsize\thepage \hfil\leftmark\mbox{}}%
\def\@oddfoot{}%
\def\@evenfoot{}}
\makeatother 
\usepackage{amsthm}
\usepackage{amsfonts}
\usepackage{bbm}
\usepackage{mathrsfs}
\usepackage{cite}
\usepackage{times}
\usepackage{graphicx}
\usepackage{amsfonts}
\usepackage{flushend}
\usepackage{algorithmic}
\usepackage[ruled]{algorithm}
\usepackage{cite}
\usepackage{times}
\usepackage{amsmath}
\usepackage{graphicx}
\usepackage{epsfig}
\usepackage{amsmath}
\usepackage{latexsym}
\usepackage{amsfonts}
\usepackage{amssymb}
\usepackage{paralist}
\usepackage{textcomp}%homepage slide valid
\usepackage{xspace}
\usepackage{mathrsfs}
\usepackage{amssymb}
\usepackage{color}
\usepackage{algorithmic}
\usepackage[ruled]{algorithm}
\usepackage{flushend}

\usepackage{supertabular}
\usepackage{subfigure}
\usepackage{subfig}
\usepackage{epstopdf}
\usepackage{multirow}
\usepackage{mdwmath}
\usepackage{booktabs}
\usepackage{tabularx}

\newtheorem{defn}{Definition~}

\renewcommand{\paragraph}[1]{\smallskip \noindent {\textsc{#1}}}

\newcommand{\mycut}[1]{{}}
\DeclareGraphicsExtensions{.eps,.ps,.pdf}
\renewcommand{\baselinestretch}{1.01}

\begin{document}
\title{Composite Behavioral Modeling for Identity Theft Detection in Online Social Networks}
\author{Cheng Wang,~\IEEEmembership{Senior Member,~IEEE},
and Bo Yang
\IEEEcompsocitemizethanks{
   \IEEEcompsocthanksitem   Wang and Yang are with the Department of Computer Science and Technology, Tongji University, and with the Key Laboratory of Embedded System and Service Computing, Ministry of Education, China.
(E-mail: chengwang@tongji.edu.cn, boyang@tongji.edu.cn)
}
}
\IEEEcompsoctitleabstractindextext{
\begin{abstract}
%\boldmath
%The profit of compromising online accounts, e.g., accounts in online social networks (OSNs) and online trading services (OTSs),
% has been attracting more cybercriminals.
%It is a necessary service guarantee to establish a non-intrusive and continuous authentication scheme to detect identity thieves.
%Behavior-based detection is convincingly an effective and then promising method to this end.
%Some existing behavior-based methods capture some deliberate content to identify suspicious accounts, e.g., utilizing malicious URLs or sensitive keywords.
%However, this type of methods are impractical in a real-life scenario where cunning criminals usually avoid using compromised accounts to send sensitive messages.
%Other existing methods can solve this problem by detecting the mutation of individual's behavior pattern.
%But the efficacy of such individual-level behavior-based methods significantly suffer from the poor quality of behavior records due to data collecting limitations or some privacy issues.
In this work, we aim at building a bridge from poor  behavioral data to an effective, quick-response, and robust behavior model for online identity theft detection.
We concentrate on this issue in online social networks (OSNs) where users usually have composite behavioral records, consisting of multi-dimensional low-quality data, e.g., offline check-ins and online user generated content (UGC).
As an insightful result, we find that there is a complementary effect among different dimensions of records for modeling users' behavioral patterns.
To deeply exploit such a complementary effect, we propose a joint model to capture both online and offline features of a user's composite behavior.
We evaluate the proposed joint model by comparing with some typical models on two real-world datasets: Foursquare and Yelp.
In the widely-used setting of theft simulation (simulating thefts via behavioral replacement), the experimental results show that our model outperforms the existing ones, with the  AUC values $0.956$ in Foursquare and $0.947$ in Yelp, respectively.
Particularly, the \emph{recall} (True Positive Rate) can reach up to $65.3\%$ in Foursquare and $72.2\%$ in Yelp with the corresponding  \emph{disturbance rate} (False Positive Rate) below $1\%$.
It is worth mentioning that these performances can be achieved by examining only one composite behavior (visiting a place and posting a tip online simultaneously) per authentication, which guarantees the low response latency of our method.
This study would give the cybersecurity community new insights into whether and how a real-time online identity authentication can be improved via modeling users' composite behavioral patterns.
\end{abstract}

}
\maketitle
\IEEEdisplaynotcompsoctitleabstractindextext
\IEEEpeerreviewmaketitle

\section{Introduction}\label{section-Introduction}
% no \IEEEPARstart

With the rapid development of the Internet, more and more affairs, e.g., mailing \cite{OnaolapoMS16}, health caring \cite{medical}, shopping \cite{de2015unique}, booking hotels and purchasing tickets are handled online \cite{hyman2013cybercrime}.
While, the Internet brings sundry potential risks of invasions \cite{lynch2005identity}, such as losing financial information \cite{bilge2009all}, identity theft \cite{Newman2005Identity} and privacy leakage \cite{de2015unique}.
Online accounts serve as the agents of users in the cyber world.
Online identity theft is a typical online crime which is the deliberate use of other person's account \cite{EgeleSKV17}, usually as a method to gain a financial advantage or obtain credit and other benefits in other person's name.
In fact, compromised accounts are usually the portals of most cybercrimes \cite{ThomasLZBRIMCEM17,OnaolapoMS16}, such as blackmail \cite{bilge2009all}, fraud \cite{pratt2010routine} and spam \cite{URLSpamFiltering}.
Thus, identity theft detection is essential to guarantee users' security in the cyber world.

Traditional identity authentication methods \cite{marshall2005identity,schneier2005two} are mostly based on access control schemes, e.g., passwords and tokens \cite{DíazSantiago2016}.
%Besides the overhead for users, some reasons make them incapable of protecting security in real-time online services \cite{waldrop2016hack}:
But users have to spend overheads in managing dedicated passwords or tokens.
Accordingly, the biometric identification \cite{CEREBRE, Labati2016BRA29662782933241,SitovaSYPZGB16,RajoubZ14} is delicately introduced to start the era of password free.
However, some disadvantages make these access control schemes still incapable of being effective in real-time online services \cite{waldrop2016hack}:
%%%%%%%%%%%%%%%%%%%%%%%%%%%%%%
(1)  They are not \emph{non-intrusive}.
Users have to spend extra time in authentication.
%%%%%%%%%%%%%%%%%%%%%%%%%%%%%%%%
% the complex passwords and so many tokens, users are easy to forget the password or leak the token,  and it is a in.
%Second and have to spend extra vigor and time passing identification.
(2)  They are not \emph{continuous}.
The defending system will  fail to take further protection once the access control is broken.

Behavior-based suspicious account detection \cite{waldrop2016hack} is a highly-anticipated solution to pursue a non-intrusive and continuous identity authentication for online services.
It depends on capturing users' suspicious behavior patterns to discriminate the suspicious accounts.
%%%%%%%%%%%%%%%%%%%%%%%%%%%%%%%%%%%%%
%Generally, there are two types of suspicious behaviors \cite{Gradon13,Shu-2017-acm,wall2007cybercrime}:
%
%
%(1) Population-Level (PL) Suspicious Behavior.
%This represents the behavioral pattern of some users that does not conform to that of the majority of users.
%Usually, in the online identity theft detection problem,
%the number of positive and negative instances is extremely imbalance.
%Then PL suspicious behaviors can be regarded as   \emph{outliers}.
%The users with such PL suspicious behaviors can be reasonably regarded as crime suspects, e.g., the suspects of fake/sybil account creators or identity thieves \cite{emigh2005online}.
%
%
%
%
%
%
%(2) Individual-Level (IL) Suspicious Behavior.
%This is an \emph{anomalous behavior}, denoting the behavioral pattern of a user that does not conform to his/her previous pattern.
%The user with such an IL suspicious behavior can be regarded as an identity theft suspect.
%%%%%%%%%%%%%%%%%%
The problem can be divided into two categories:  fake/sybil account detection and  compromised account detection \cite{mercuri2006scoping}.
The fake/sybil account's behavior usually does not conform to the behavioral pattern of the majority.
%Usually, in the online identity theft detection problem,
%the number of normal and abnormal users is extremely imbalance.
%Then PL suspicious behaviors can be regarded as   \emph{outliers}.
%The users with such PL suspicious behaviors can be reasonably regarded as crime suspects, e.g., the suspects of fake/sybil account creators or identity thieves \cite{emigh2005online}.
While, the compromised account usually behaves in a pattern that does not conform to his/her previous one, even behaves like fake/sybil accounts.
It can be solved by capturing \emph{mutations} of users' behavioral patterns.

%After the corresponding accounts are compromised, identity thieves' behaviors usually mutate to either an \emph{outlier} or \emph{anomalous}  pattern.
%Accordingly, these mutations can be captured orderly by the modules of detecting  outlier  and  anomalous suspicious behaviors.
%% as the procedure depicted in Fig. \ref{behavior-types-frame}.

Comparing with detecting compromised accounts, detecting fake/sybil accounts is relatively easy, since the latter's behaviors are generally more detectable than the former's.
It has been extensively studied, and can be realized by various population-level approaches, e.g., clustering \cite{EVILCOHORT,UncoveringMaliciousAccounts}, classification \cite{stringhini2010detecting,bilge2009all} and statistical or empirical rules \cite{ahmed2013generic,pratt2010routine,milne2009toward,abbasi2010detecting}.
%%%%%%%%%%%%%%%%%%%%%%%%%%%%%%%%%%%%%%%%%%%%%%%%%%%%%%%%%%%%%%%%%%%%%%%%%%%%%%%%%%%%%%%%%%%%%%%%%%%%
Thus, we \emph{only} focus on the compromised account detection, commonly-called \emph{identity theft detection}, based on individual-level behavior models.

Recently, researchers have proposed different individual-level identity theft detection methods based on  suspicious behavior detection \cite{Keystroke,AbouelenienPMB17,LesaegeSLV16,ProfilingOnlineSocialBehaviors,NokhbehZaeem201750, URLSpamFiltering,WarningBird,POIFriends}.
However, the efficacy of these methods significantly depends on the sufficiency of behavior records, suffering from the low-quality of behavior records due to data collecting limitations or some privacy issues \cite{de2015unique}.
Especially, when a method only utilizes a specific dimension of behavioral data, the efficacy damaged by poor data is possibly enlarged.
Unfortunately, most existing works just concentrate on a specific dimension of users' behavior, such as keystroke \cite{Keystroke}, clickstream \cite{ProfilingOnlineSocialBehaviors}, and user generated content (UGC) \cite{NokhbehZaeem201750, URLSpamFiltering,WarningBird}.

%%%%%%%%%%%%%%%%%%%%%%%%%%%%%%%%%%%%%%%%%%%%%%%%%%%%%%%%%%%%%%%%%%%%%%%%
%\begin{figure}[t]
% %\renewcommand{\captionlabelfont}{\footnotesize}
%% \captionsetup{font={footnotesize}}
%  \centering
%    \includegraphics[width=0.47 \textwidth]{frame.pdf}
%   % \includegraphics[width=0.45\textwidth]{imagesrvice-latency-180.eps}}
%  \caption{Suspicious Account Detection (SAD). The SAD problem is generally divided into two types: Fake/Sybil Account Detection and Compromised Account Detection. The latter is the focus of this work, i.e., so-called identity theft detection.
% Behavior-based methods are applicable to   both ones.}
%  \label{behavior-types-frame} %% label for second subfigure
%\end{figure}
%%%%%%%%%%%%%%%%%%%%%%%%%%%%%%%%%%%%%%%%%%%%%%%%%%%%%%%%%%%%%%%%%%%%%%%%

\begin{figure}[t]
 %\renewcommand{\captionlabelfont}{\footnotesize}
% \captionsetup{font={footnotesize}}
  \centering
    \includegraphics[width=0.45 \textwidth]{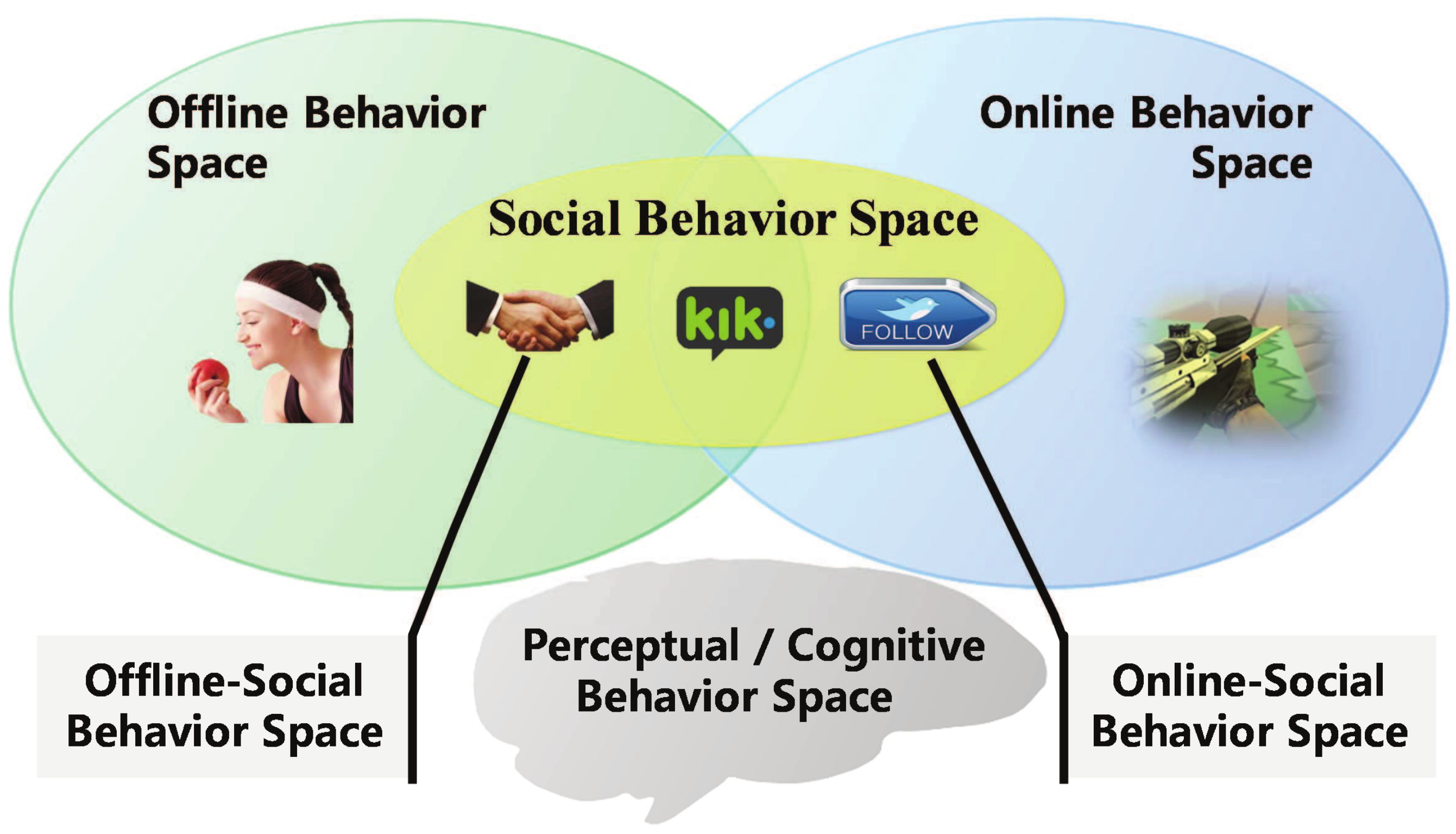}
 \vspace{-0.1in}
  \caption{An illustration of composite behavior space.}
  \label{behaviorspace} %% label for second subfigure
\vspace{-0.15in}
\end{figure}

In this paper, we propose an approach to detect identity theft by jointly using multi-dimensional behavior records which are possibly insufficient in each dimension.
According to such characteristics, we choose the online social network (OSN) as a typical scenario where most users' behaviors are coarsely recorded \cite{Wangsigir17}.
%, and OSN users are indeed likely to encounter identity thieves\footnote{The largest privately held cybersecurity organization based in the USA, https://www.webroot.com/}.
In the Internet era, users' behaviors are composited by offline behaviors, online behaviors, social behaviors, and perceptual/cognitive behaviors, as illustrated in Fig. \ref{behaviorspace}.
For OSN users, the behaviorial data are collectable in many daily life applications, such as offline check-ins in location-based services, online tips-posting in UGC sites, and social relationship-making in OSN sites \cite{Foursquare,YinHZWZHS16,POIFriends}.
%%%%%%%%%%%%%%%%%%%%%%%%%%%%%%%%%%%%%%%%%
Accordingly, we design our method based on users' composite behaviors by these categories shown in Fig. \ref{behaviorspace}.

%%%%%%%%%%%%%%%%%%%%%%%%%%%%%%%%%%%%%%%%%

%%%%%%%%%%%%%%%%%%%%%%%%%%%%%%%%%%%%%%%%%%
%we mainly devise two critical techniques as follows:
%%%%%%%%%%%%%%%%%%%%%%%%%%%%%%%%%%%%%%%%%%%%%%%
%(1) \emph{Crossed Data Filling Method}.
%In real-life scenario, most users have only several behavioral records that are too insufficient to build qualified behavior models.
%The urgent priority is to improve data quality by transforming data.
%We develop a data filling scheme, called \emph{crossed data filling}. Under the scheme,
%the similarities among users' visiting patterns
%are first exploited to fill check-in data by collaborative filtering,
%and then for every user, the social relationship is utilized to fill his/her tips-posting data on every visited venue.
%Compared with the widely-used tensor decomposition \cite{Li-2017-acm-tois} and embedding \cite{Xie-2016-cikm} methods, our filling method serves as a lightweight data filling scheme to obtain sufficient users' records for behavior modeling.
%%%%%%%%%%%%%%%%%%%%%%%%%%%%%%%%%%%%%%%%%%%%%%%%%%%%%
%
%With an intuition that a user is likely to have similar interests with her friends.
%We can use friendships in OSNs to improve data quality and assume that each user has a chance $\varepsilon$ to do like her friends.
%However, different friends have different impact on the user.
%Thus, we utilize collaborative filtering to measure user-friend similarity and use it to represent their potential impact.

We devote to proving that a high-quality (effective, quick-response, and robust) behavior model can be obtained by integrally using multi-dimensional behaviorial data, even though the quality of data is extremely insufficient in each dimension.
For this challenging objective, a precondition is to solve users' data insufficiency problem.
The majority of users commonly have only several behavioral records that are too insufficient to build qualified behavior models.
For this issue, we adopt a tensor decomposition-based method \cite{WangZX14} by combining the similarity among users (in terms of interests to both tips and places) with social ties among them.
 %(2) \emph{Joint Probabilistic Generative Model}.
Then, to fully utilize potential information in composite behaviors for user profiling, we propose a joint probabilistic generative model based on Bayesian networks, called \emph{Composite Behavioral Model} (CBM).
It offers a composition of the typical features in two different behavior spaces: check-in location in offline behavior space and UGC in online behavior space.
%%%%%%%%%%%%%%%%%%%%%%%%%%%%%%%%%%%%%%%%%%%%%%%%%%%%%%%
%summary of observations
Consider a composite behavior of a user, we assume that its generative mechanism is as follows:
When a user plans to visit a venue and simultaneously post tips online, he/she subconsciously select a specific behavioral pattern according to his/her behavior distribution.
Then, he/she comes up with a topic and a targeted venue based on the present pattern's topic and venue distributions, respectively.
Finally, his/her comment words are generated following the corresponding topic-word distribution.
%Consider a specific composite behavior of a user $u$, i.e., visiting a venue $v$ and simultaneously posting online a tip on the topic $z$.
%We assume that the underlying generative mechanism of the behavior is as follows:
%User $u$ plays multiple ``roles" in daily life.
%Before the specific behavior, he first chooses a \emph{role}, described by a corresponding \emph{community} $c$,
%according to his community distribution $\pi_u$.
%Then, he  selects a venue $v$ and a topic $z$ according to the venue and topic distributions depending on the community/role $c$, denoted by $\vartheta_c$ and $\theta_c$, respectively.
%Finally, an online tip $\mathcal{D}$ is generated under the topic $z$. That is, each word $w$ in the message is generated by the corresponding word distribution $\phi_{z,w}$.
%下面这句是修改还是注释掉？
To estimate the parameters of the mentioned distributions, we adopt the collapsed Gibbs sampling \cite{YinHZWZHS16}.

Based on the joint model CBM, for each composite behavior, denoted by a triple-tuple $(u, v, \mathcal{D})$, we can calculate the chance of user $u$ visiting venue $v$ and posting a tip online with a set of words $\mathcal{D}$.
Taking into account different levels of activity of different users, we devise a \emph{relative anomalous score} $S_r$ to measure the occurrence rate of each composite behavior $(u, v, \mathcal{D})$.
By these approaches, we finally realize a real-time (i.e., judging by only one composite behavior) detection for identity theft suspects.

%An intuition is that a with lower occurrence rate can be regarded as anomalous behavior.
%However, the experiment results indicate that \emph{logarithmic anomalous score} ($S_l$) \emph{cannot} detect identity theft cases well.
%A possible explanation for the \emph{counterintuitive} result is that .
%Consequently, we adopt

We evaluate the proposed joint model by comparing with the typical models on two real-world OSN datasets: Foursquare \cite{Foursquare} and Yelp \cite{Yelp}.
We adopt the \emph{area under the receiver operating characteristic curve} (AUC) as the detection efficacy.
In the widely-used setting of theft simulation (simulating thefts via behavioral replacement) \cite{LichmanS14}, the AUC value reaches $0.956$ in Foursquare and $0.947$ in Yelp, respectively.
Particularly, the \emph{recall} (True Positive Rate) reaches up to $65.3\%$ in Foursquare and $72.2\%$ in Yelp with the corresponding  \emph{disturbance rate} (False Positive Rate) below $1\%$.
Note that these performances can be achieved by examining only one composite behavior per authentication, which guarantees the low response latency of our detection method.
%%%%%%%%%%%%%%%%%%%%%%%%
As an insightful result, we learn that the complementary effect does exist among different dimensions of low-quality records for modeling users' behaviors.
To the best of our knowledge, this is the first work jointly leveraging online and offline behaviors to detect identity theft in OSNs.

The rest of this paper is organized as follows.
%Firstly, we present the related work in Section \ref{section-Related}.
We give an overview of our solution in Section \ref{section-problel-soulution}.
Then, we present our method in Section \ref{section-method}, and make the validation in Section \ref{section-evaltuationj}.
We provide a literature review in Section \ref{section-Related}.
Finally, we draw conclusions in Section \ref{section-Conclusion}.

%\vspace{-0.05in}
\section{Overview of Our Solution}\label{section-problel-soulution}

%In this section, we
%introduce
% the issue of online identity theft detection and
 %give an overview of our proposed method.
%\subsection{Online Identity Theft Detection}
%It is usually easier to deceive others in the Internet compared to the real world due to some factors such as the ease of leaking identity and the perception of lower moral cost.
%People using OSNs tend to have more online friends than real-life friends.
%Usually, it is hard to verify their online friends' identities but people are still willing to share their personal information even they know there are privacy issues in such online world.

Online identity theft occurs when a thief steals a user's personal data and impersonates the user's account.
%In Fig. \ref{IDT}, we present a typical behavior flow of online identity theft cases.
Generally, a thief usually first gathers information about a targeted user to steal his/her identity and then use the stolen identity to interact with other people to get further benefits \cite{ThomasLZBRIMCEM17}.
%On the other hand, victim's friends may notice and report the suspicious behaviors to help security center catches the thief.
%%%%%%%%%%%%%%%%%%%%%%%%%
Criminals in different online services usually have different motivations.
%
%In some OSNs where one may know their online friends' real-life identities,
%thieves usually utilize the strong trust between friends to obtain benefit.
%These behaviors usually contain specific sensitive words.
%In other scenarios where friends may not familiar with each other in real-life and lack directly interactions (i.e., real-time communications),
%thieves can not obtain direct benefit from cheating users' friends, then turn to spread malicious messages.
%Some malicious messages have explicit features such as URLs and contact numbers.
%But some malicious messages only contain deceptive comments to a place, an actor or an event, which look like normal comments and seems harder to detect.
%%%%%%%%%%%%%%%%%%%%%%%%%%%%%%%%%%%%%%%%%%%%%%%%%%%%%%%%%%%%%%%%%%%%%%%%%%%%%%%%%%
%\begin{figure}
% %\renewcommand{\captionlabelfont}{\footnotesize}
%% \captionsetup{font={footnotesize}}
%  \centering
%    \includegraphics[width=0.4 \textwidth]{identity-theft-self.pdf}
%   % \includegraphics[width=0.45\textwidth]{imagesrvice-latency-180.eps}}
%  \caption{The flow chart of typical attack and defense policies in identity theft cases.}
%  \label{IDT} %% label for second subfigure
%\end{figure}
%%%%%%%%%%%%%%%%%%%%%%%%%%%%%%%%%%%%%%%%%%%%%%%%%%%%%%%%%%%%%%%%%%%%%%%%%%%%%%%%%%
%\subsection{Composite Behavioral Model for Identity Identification}
%Traditional identity identification methods usually focus on explicit features.
%However, there are many cases have no explicit features.
In this work, we focus on online social networks (OSNs).
In some OSNs, one may know their online friends' real-life identities.
Thieves usually utilize the strong trust between friends to obtain benefit \cite{BurszteinBMPAAPS14}.
Their behavioral records usually contain specific sensitive terms.
In other scenarios, friends may not be familiar with each other and lack direct interactions in the real world, which makes thieves can not obtain direct benefit from cheating users' friends.
So they turn to spread malicious messages in these OSNs.
Among these malicious messages, some have explicit features such as URLs and contact numbers, others only contain deceptive comments on a place, a star or an event.
The latter messages look like normal ones, which makes them harder to detect.
Thus, we apply a widely-used setting of theft simulation, i.e., simulating thefts via behavioral replacement, to represent this kind of thieves.

An OSN user's behavior is usually composite of online and offline behaviors occurring in different behavioral spaces, as illustrated in Fig. \ref{behaviorspace}.
Based on this fact, we aim to propose a joint model to embrace them into a unified model to deeply extract information.
% from a specific composite behavior.
%we propose two composite behavioral models, which combining sub-behaviors in two different ways.
%Our fused model directly combines CF-KDE and LDA to use both offline and online behavioral features.
%The detail is provided in Section \ref{Experiment-analysis}.
%A shortage in the fused model is that it does not consider the correlations between offline and online behaviors.
%Thus, we propose a joint model to embrace them in a unified model to describe a user/behavior.

Before introducing our joint model, named \emph{Composite Behavioral Model} (CBM), we provide some conceptions as preparations.
The relevant notations are listed in Table \ref{notation}.
%All vectors are column vectors and are denoted by bold lower case letters (e.g., $\mathbf{\theta}$ and $\mathbf{\phi}$).
%We use calligraphic letters to represent sets (e.g., $\mathcal{U}$ and $\mathcal{V}$).
%Some important are summarized as follows.

\begin{table}[t]\renewcommand{\arraystretch}{1.4}
  \centering
  \caption{Notations of Parameters}\label{notation}
 % \vspace{-0.1in}
  \scalebox{1.0}{\begin{tabular}{p{1.1cm}|p{6.9cm}}
  \toprule
   \ Variable&Description\\
  \hline
   \hline
   \ $w$&the word in UGC\\
  \hline
  \ $v$&the venue or place\\
  \hline
  \ $\pi_u$&the community memberships of user $u$, expressed  by a multinomial distribution over communities\\
  \hline
  \ $\theta_c$&the interests of community $c$, expressed by   a multinomial distribution over topics\\
  \hline
  \ $\vartheta_c$&a multinomial distribution over spatial items   specific to community $c$\\
  \hline
  \ $\phi_z$&a multinomial distribution over words   specific to topic $z$\\
  \hline
%  \ $h_{c,v}$&the bandwidth vector of the kernel function \\ & specific to $(c, v)$\\
%  \hline
$\alpha, \beta, \gamma, \eta$&Dirichlet priors to multinomial distributions   $\theta_c$, $\phi_z$, $\pi_u$ and $\vartheta_c$, respectively\\
  \bottomrule
  \end{tabular}
  }
%  \vspace{-0.22in}
\end{table}

%\vspace{-0.08in}
\begin{defn} [Composite Behavior]
A composite behavior, denoted by a four-tuple $(u, v, \mathcal{D}, t)$, indicates that at time $t$, user $u$  visits venue $v$ and simultaneously posts online a tip consisting of a set of words $\mathcal{D}$.
% which are extracted from all tips of users and descriptions of venues.
\end{defn}

%Let $\mathcal{B}_u$ denote a collection of user $u$'s behaviors.

We remark that for a composite behavior, the occurring time $t$ is a significant factor.
Two types of \emph{time attributes} play important roles in digging potential information for improving the identification.
The first is the \emph{sequential correlation} of behaviors.
However, in real-life OSNs, the time intervals between adjacent recorded behaviors are mostly unknown or overlong, which leads that the temporal correlations cannot be captured effectively.
The second  is the \emph{temporal property} of behaviors, e.g., periodicity and preference variance over time.
However, in some real-life OSN datasets, the occurring time is recorded with a low resolution, e.g., by day, which shields the possible dependency of a user's behavior on the occurring time.
Thus, it is difficult to obtain reliable time-related features of users' behaviors.
Since we aim to propose a practical method based on uncustomized datasets of user behaviors, we only concentrate on the dependency between a user's check-in location and tip-posting content of each behavior, taking no account of the impact of specific occurring time in this work.
Thus, the representation of a composite behavior can be simplified into a triple-tuple $(u, v, \mathcal{D})$ without confusion in this paper.

%%%%之所以不加 t，是因为时间t的两大作用均失效：第一，行为时间序列相关性，往往体现在相邻的行为之间，但是现实osn用户行为记录并不是连续的，间隔很大，这种行为之间的相关性是很弱的，甚至是对结果误导性的。第二，其他维度行为与时间本身的相关性，比如，中午去餐馆比较正常，但是，现实osn用户行为记录中的时间记录往往是resolution太小，以天为单位，这个掩盖了很多的相关性，从而，失效。在这个工作中，我们也并没有考虑进去。

Our model depends on the following assumptions:
(1) Each user behaves in multiple patterns with different possibilities;
(2) Some users  have similar behavioral patterns, e.g., similar interests in topics and places.

To describe the features of  users' behaviors, we first introduce the \emph{topic} of tips.

%\vspace{-0.02in}
\begin{defn}[Topic, \cite{Blei2003LDA944919944937}]
Given a set of words $\mathcal{W}$, a topic $z$ is represented by a multinomial distribution over words, denoted by $\mathbf{\phi}_z$, whose every component $\phi_{z,w}$ denotes the possibility of word $w$ occurring in topic $z$.
\end{defn}

Next, we formulate a specific \emph{behavioral pattern} of users by a conception called \emph{community}.

%\vspace{-0.03in}
\begin{defn} [Community]\label{defn-comunity}
A community is a set of users with a similar behavioral pattern.
Let $\mathcal{C}$ denote the set of all communities.
A community $c \in \mathcal{C}$ has two critical parameters:
(1) A topic distribution $\mathbf{\theta}_c$ whose component, say $\theta_{c,z}$, indicates the probability that users in community $c$ send a message with topic $z$.
(2) A spatial distribution $\mathbf{\vartheta}_c$ whose component, say $\vartheta_{c,v}$, represents the chance that users in community $c$ visit venue $v$.
\end{defn}

%In our model, a community $c \in \mathcal{C}$ determines two parameters:
%1) a multinomial distribution over topics $\mathbf{\theta}_c$, each component $\theta_{c,z}$ indicates the probability that users in community $c$ send a message with topic $z$;
%2) a multinomial distribution over venues $\mathbf{\vartheta}_c$, each component $\vartheta_{c,v}$ represents the chance that users in community $c$ visit venue $v$.

%%%%将community改成一个role pattern.%%%%%%%%%%%%%

%A user $u$'s behavior is usually reasonable.
%For instance, people are likely to work in the company and watch TV at home.
%We can see that people are in two different communities (play two roles in office and home) for these two behaviors.
%Thus, we believe that the behavior is determined by the user's current community/role.

More specifically, we assume that a community is formed by the following procedure:
Each user $u$ is included in communities according to a multinomial distribution, denoted by $\mathbf{\pi}_u$.
That is, each component of $\mathbf{\pi}_u$, say $\pi_{u,c}$, denotes $u$'s affiliation degree to community $c$.
Similarly, we allocate each community $c$ with a topic distribution $\mathbf{\theta}_c$ to represent its online topic preference and a spatial distribution $\mathbf{\vartheta}_c$ to represent its offline mobility pattern.

%To detect identity theft cases, we propose a joint probabilistic generative model based on Bayesian network.
%The graphical representation of our joint model is demonstrated in Figure \ref{Graphical-representation}.
%The notations of the model are listed in Table \ref{notation}.
\begin{figure}
\centering
    \includegraphics[width=0.45 \textwidth]{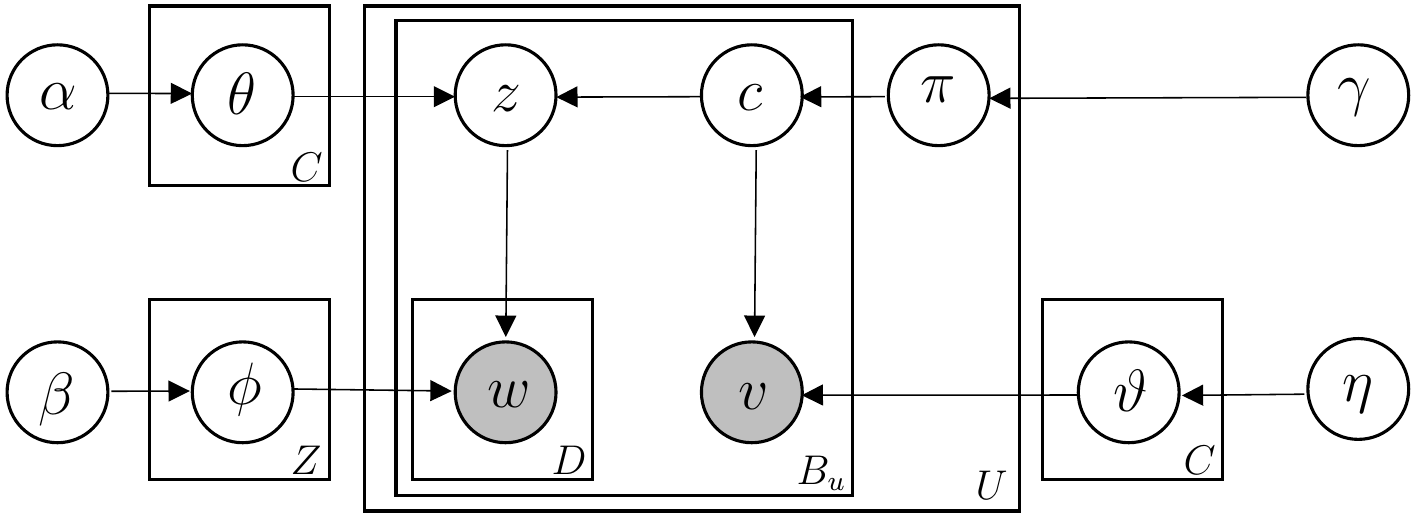}
  \caption{Graphical representation of joint model. The parameters are explained in Table \ref{notation}.}
  \label{Graphical-representation}
 % \vspace{-0.1in}
\end{figure}

%Our joint model depends on
%(1) each user has different communities/roles in his/her real-life;
%(2) each user plays one role in each behavior;
%(3) users within the same community/role have similar chooses (i.e., where to go);
%(4) each behavior indicates one topic.

The graphical representation of our joint model  CBM is demonstrated in Fig. \ref{Graphical-representation}.

\section{Method}\label{section-method}

\subsection{Composite Behavioral Model}

Generally, users take actions according to their regular behavioral patterns which are represented by the corresponding communities (Definition \ref{defn-comunity}).
We present the behavioral generative process in Algorithm \ref{generative-process}:
When a user $u$ is going to visit a venue and post online tips there, he/she subconsciously select a specific behavioral pattern, denoted by community $c$, according to his/her community distribution $\mathbf{\pi}_u$ (Line $11$).
Then, he/she comes up with a topic $z$ and a targeted venue $v$ based on the present community's topic and venue distributions ($\mathbf{\theta}_c$ and $\mathbf{\vartheta}_c$, respectively) (Line $12-13$).
Finally, the words of his/her tips in $\mathcal{D}$ are generated following the topic-word distribution $\mathbf{\phi}_z$ (Line $15$).
%We can obtain a joint distribution of all variables in Eq. \ref{joint-distribution}.

%\vspace{-0.1in}
\begin{algorithm}[t] %算法开始
\caption{Joint Probabilistic Generative Process} %算法的题目
\label{generative-process} %算法的标签
\begin{algorithmic}[1] %此处的[1]控制一下算法中的每句前面都有标号
%\REQUIRE Text:Today is a good day. Variables:$u,v,w$. $G=(V,E)$ %输入条件(此处的REQUIRE默认关键字为Require，在上面已自定义为Input)
%\ENSURE Something... %输出结果(此处的ENSURE默认关键字为Ensure在上面已自定义为Output)
\FOR {\emph{each community} $c \in \mathcal{C}$}
\STATE Sample the distribution over topics $\mathbf{\theta}_c \thicksim Dirichlet(\cdot|\alpha)$
\STATE Sample the distribution over venues \\ $\mathbf{\vartheta}_c \thicksim Dirichlet(\cdot|\eta)$
\ENDFOR
\FOR {\emph{each topic} $z \in \mathcal{Z}$}
\STATE Sample the distribution over words \\ $\mathbf{\phi}_z \thicksim Dirichlet(\cdot|\beta)$
\ENDFOR
\FOR {\emph{each user} $u \in \mathcal{U}$}
\STATE Sample the distribution over communities \\ $\mathbf{\pi}_u \thicksim Dirichlet(\cdot|\gamma)$
\FOR {\emph{each composite behavior} $(u, v, \mathcal{D}) \in \mathcal{B}_u$}
\STATE Sample a community indicator $c \thicksim Multi(\mathbf{\pi}_u)$
\STATE Sample a topic indicator $z \thicksim Multi(\mathbf{\theta}_c)$
\STATE Sample a venue $v \thicksim Multi(\mathbf{\vartheta}_c)$
\FOR {\emph{each word} $w \in \mathcal{D}$}
\STATE Sample a word $w \thicksim Multi(\mathbf{\phi}_z)$
\ENDFOR
\ENDFOR
\ENDFOR
\end{algorithmic}
\end{algorithm}

Exact inference of our joint model CBM is difficult due to the intractable normalizing constant of the posterior distribution, \cite{YinHZWZHS16}.
We adopt collapsed Gibbs sampling for approximately estimating distributions (i.e., $\mathbf{\theta}$, $\mathbf{\vartheta}$, $\mathbf{\phi}$ and $\mathbf{\pi}$).
As for the hyperparameters, we take a fixed value, i.e., $\alpha=50/Z$, $\gamma=50/C$ and $\beta=\eta=0.01$, following the study in \cite{HuYCX15}, where $Z$ and $C$ are the numbers of topics and communities, respectively.

In each iteration, for each composite behavior $(u, v, \mathcal{D})$, we first sample community $c$ according to Eq. (\ref{sample-c}):
\begin{equation}\label{sample-c}
P(c|\mathbf{c^{\neg}}, \mathbf{z}, \mathbf{v}, u) \propto (n_{u,c}^{\neg}+\gamma) \frac{n_{c,z}^{\neg}+\alpha}{\sum_{z'}(n_{c,z'}^{\neg}+\alpha)} \frac{n_{c,v}^{\neg}+\eta}{\sum_{v'}(n_{c,v'}^{\neg}+\eta)},
\end{equation}
where $\mathbf{c^{\neg}}$ denotes the community allocation for all composite behaviors except the current one;
$\mathbf{z}$ denotes the topic allocation for all composite behaviors;
$n_{u,c}$ denotes the number of times that community $c$ is generated by user $u$;
$n_{c,z}$ denotes the number of times that topic $z$ is generated by community $c$;
$n_{c,v}$ denotes the number of times that venue $v$ is visited by users in community $c$;
a superscript $\neg$ denotes something except the current one.

Then, given a community $c$, we sample topic $z$   according to the following Eq. (\ref{sample-z}):
\begin{equation}\label{sample-z}
P(z|\mathbf{z^{\neg}}, \mathbf{c}, \mathcal{D}) \propto (n_{c,z}^{\neg}+\alpha) \prod_{w \in \mathcal{D}}\frac{n_{z,w}^{\neg}+\beta}{\sum_{w'}(n_{z,w'}^{\neg}+\beta)},
\end{equation}
where $n_{z,w}$ denotes the number of times that word $w$ is generated by topic $z$.
%After $I_b$ iterations, we start to estimate parameters every $I_s$ iterations,
%where $I_b$ and $I_s$ are determined in Algorithm \ref{inference}.

The inference algorithm is presented in Algorithm \ref{inference}.
We first randomly initialize the topic and community assignments for each composite behavior (Line  $2-4$).
Then, we update the community and topic assignments for each composite behavior based on Eqs.  (\ref{sample-c}) and (\ref{sample-z}) in each iteration (Line  $6-9$).
Finally, we estimate the parameters, test the coming cases and update the training set every $I_s$ iterations since $I_b$th iteration (Line $10-13$) to address concept drift.

\begin{algorithm}[t] %算法开始
\caption{Inference Algorithm of the Joint Model CBM} %算法的题目
\label{inference} %算法的标签
\begin{algorithmic}[1] %此处的[1]控制一下算法中的每句前面都有标号
\REQUIRE user composite behavior collection $\mathcal{B}$, number of iteration $I$, start saving step $I_b$, saving lag $I_s$, start training sequence number $N_b$, end training sequence number $N_e$, hyperparameters $\alpha$, $\beta$, $\gamma$ and $\eta$
\ENSURE estimated parameters $\hat{\mathbf{\theta}}$, $\hat{\mathbf{\vartheta}}$, $\hat{\mathbf{\phi}}$, $\hat{\mathbf{\pi}}$
\STATE Create temporary variables $\mathbf{\theta^{sum}}$, $\mathbf{\vartheta^{sum}}$, $\mathbf{\phi^{sum}}$ and $\mathbf{\pi^{sum}}$, initialize them with zero, set testing sequence number $N_t=0$ and let $\mathcal{B}(N_t)$ denotes the corresponding training collection for testing behaviors which sequence number values $N_t$
\FOR {\emph{each composite behavior $(u, v, \mathcal{D}) \in \mathcal{B}(N_t)$}}
\STATE Sample community and topic randomly
\ENDFOR
%\STATE Initialize variable \emph{count} with zero
\FOR {$iteration=1$ to $I$}
\FOR {\emph{each behavior $(u, v, \mathcal{D}) \in \mathcal{B}(N_t)$}}
\STATE Sample community $c$ according to Eq. (\ref{sample-c})
\STATE Sample topic $z$ according to Eq. (\ref{sample-z})
\ENDFOR
\IF {$(iteration>I_b)$ and $(iteration~mod~ I_s==0)$}
\RETURN model parameters as follows: \\
$\theta_{c,z} = \frac {n_{c,z}+\alpha} {\sum_{z'}(n_{c,z'}+\alpha)}$;
$~~\vartheta_{c,v} = \frac {n_{c,v}+\eta} {\sum_{v'}(n_{c,v'}+\eta)}$ \\
$\pi_{u,c} = \frac {n_{u,c}+\gamma} {\sum_{c'}(n_{u,c'}+\gamma)}$;
$~~\phi_{z,w} = \frac {n_{z,w}+\beta} {\sum_{w'}(n_{z,w'}+\beta)}$
\STATE Evaluate corresponding test cases and update $N_t++$; $N_b++$; $N_e++$
\ENDIF
\ENDFOR
%\RETURN model parameters \\$\hat{\mathbf{\theta}}=\frac{\mathbf{\theta^{sum}}}{count}$, $\hat{\mathbf{\vartheta}}=\frac{\mathbf{\vartheta^{sum}}}{count}$, $\hat{\mathbf{\phi}}=\frac{\mathbf{\phi^{sum}}}{count}$ and $\hat{\mathbf{\pi}}=\frac{\mathbf{\pi^{sum}}}{count}$
\end{algorithmic}
\end{algorithm}

%%%%%%%%%%%%%%%%%%%%%%%%%%%%%%%%%%%%%%%%%%%%%%%%%%%%%
%To overcome the problem of data sparsity, we take users' friendships into consideration.
%%For each user $u$, we assume that user $u$ has a chance $\varepsilon$ to do the same behavior like his/her friends.
%%In our work, we let $\varepsilon=0.01$.
%With an argument that a user is likely to have similar interests with her friends \cite{LesaegeSLV16, HuangZSWH17},
%we can utilize friendships in OSNs to improve data quality, and assume that each user has a chance $\varepsilon$ to do like her friends.
%However, different friends have different impact on the user.
%Thus, we furthermore introduce the method of collaborative filtering \cite{ChenZ0NLC17}, as illustrated in Section \ref{subsub-mehodes-repreign}, to measure the user-friend similarity, and use it to represent their potential impact.
%%%%%%%%%%%%%%%%%%%%%%%%%%%%%%%%%%%%%%%%%%%%%%%%%%%
%data filling
To overcome the problem of data insufficiency, we adopt the tensor decomposition \cite{WangZX14} to discover their potential behaviors.
In our experiment, we use the Twitter-LDA \cite{ZhaoJWHLYL11} to obtain each UGC's topic and construct a tensor $\mathbf{A} \in \mathbb{R}^{N \times M \times L}$, with three dimensions standing for users, venues and topics.
Then, $\mathbf{A}(u,v,z)$ denotes the frequency that user $u$ posting a message on topic $z$ in venue $v$.
We can decompose $\mathbf{A}$ into the multiplication of a core tensor $\mathbf{S} \in \mathbb{R}^{d_U \times d_V \times d_Z}$ and three matrices, $\mathbf{U} \in \mathbb{R}^{N \times d_U}$, $\mathbf{V} \in \mathbb{R}^{M \times d_V}$, and $\mathbf{Z} \in \mathbb{R}^{L \times d_Z}$, if using a tucker decomposition model, where $d_U$, $d_V$ and $d_Z$ denote the number of latent factors; $N$, $M$ and $L$ denote the number of users, venues and topics.
An objective function  to control the errors is defined as:
\begin{equation*}
\begin{split}
&\mathcal{L}(\mathbf{S},\mathbf{U},\mathbf{V},\mathbf{Z}) = \frac{1}{2} {\| \mathbf{A}-\mathbf{S} \times_{U} \mathbf{U} \times_{V} \mathbf{V} \times_{Z} \mathbf{Z} \|}^{2} \\ &+\frac{\lambda}{2}\left({\|\mathbf{S}\|}^2+{\|\mathbf{U}\|}^2+{\|\mathbf{V}\|}^2+{\|\mathbf{Z}\|}^2+\sum\nolimits_{(i,j) \in \mathcal{F}} u_{i}^{T}u_{j}\right), \label{TDM}
\end{split}
\end{equation*}
where $\mathcal{F}$ is a set of friend pairs $(i,j)$.
$\mathbf{A}^*=\mathbf{S} \times_{U} \mathbf{U} \times_{V} \mathbf{V} \times_{Z} \mathbf{Z}$ is the potential frequency tensor, and $\mathbf{A}^{*}(u,v,z)$ denotes the frequency that user $u$ may post a message on topic $z$ in venue $v$.
A higher $\mathbf{A}^{*}(u,v,z)$ indicates that user $u$ has a higher chance to do this kind of behavior in the future.
We limit the competition space to the behavior space of $u$'s friends, i.e.,
\[\left\{(u,v,z)|\mathbf{A}(u^{'},v,z)>0,(u,u^{'})\in \mathcal{F}\right\},\]
and select the top $20$ behaviors as his/her latent behavior to improve data quality.
%We assume that user $u$ behave in the same way to these similar users (i.e., if $u$ and $i$ are similar users, each behavior $(i,v,\mathcal{D})$ indicates a latent behavior $(u,v,\mathcal{D})$).

\subsection{Identity Theft Detection Scheme}
By the parameters $\hat{\mathbf{\Psi}}=\{\hat{\mathbf{\theta}}, \hat{\mathbf{\vartheta}}, \hat{\mathbf{\phi}}, \hat{\mathbf{\pi}}\}$ learnt from the inference algorithm (Algorithm \ref{inference}), we can estimate the \emph{logarithmic anomalous score} $(S_l)$ of a composite behavior $(u, v, \mathcal{D})$ by Eq. (\ref{BR}):
\begin{equation}\label{BR}
\begin{split}
S_l(u, v, \mathcal{D})&=-\lg P(v, \mathcal{D}|u)\\ &= -\lg \left(\sum_c{\hat{\pi}_{u,c} \hat{\vartheta}_{c,v} \sum_z{\hat{\theta}_{c,z}(\prod_{w \in \mathcal{D}}\hat{\phi}_{z,w})^{\frac{1}{|\mathcal{D}|}}}} \right).
\end{split}
\end{equation}

However, we may mistake some normal behaviors occurring with low probability, e.g., the normal behaviors of users whose behavioral diversity and entropy are both high, for suspicious behaviors.
Thus, we propose a \emph{relative anomalous score} $(S_r)$ to indicate the trust level of each behavior by Eq. (\ref{RBR}):
\begin{equation}\label{RBR}
S_r(u, v, \mathcal{D}) =1-P(u| v, \mathcal{D}) = 1-\frac{P(v, \mathcal{D}|u)P(u)}{\sum_{u'}P(v, \mathcal{D}|u')P(u')}.
\end{equation}
We randomly select $40$ users to estimate the relative anomalous score for each composite behavior.
Our experimental results in Section \ref{section-evaltuationj} show that the approach based on $S_r$ outperforms the approach based on $S_l$.

\section{Evaluation}\label{section-evaltuationj}
In this section, we present the experimental results to evaluate the proposed joint model CBM, and validate the efficacy of the joint model for identity theft detection on real-world OSN datasets.
\subsection{Datasets}
Our experiments are conducted on two real-life large OSN datasets: Foursquare \cite{Foursquare} and Yelp \cite{Yelp}.
\begin{table}[t]\renewcommand{\arraystretch}{1.4}
\centering
\caption{Statistics of Foursquare and Yelp Datasets}\label{t1}
%\vspace{-0.1in}
\scalebox{0.9}{\begin{tabular}{p{2.5cm}| p{2.8cm} p{2.8cm}}
% \diagbox{Height}{Group}{BMI}&\(<18\)&\(18\sim22\)&\(>22\)\\
\toprule
 \ &Foursquare&Yelp  \\
 \hline
\ \# of users&31,493&80,592\\
\ \# of venue&143,923&42,051\\
\ \# of check-ins&267,319&491,393\\
%\ \# of social ties&330,898&1,262,659\\
\bottomrule
\end{tabular}
}
\end{table}
They are two well-known online social networking service providers.
%The dataset used in our experiment contains the check-in history of $31,494$ users in LA.
%Yelp is another popular location-based social network service provider, which publishes crowd-sourced reviews about local businesses.
%The dataset used in our experiment contains the tips of $80,593$ users.
In both datasets, there is no URLs or other sensitive terms.
Both datasets contain users' social ties and behavioral records.
Each social tie contains \emph{user-ID} and \emph{friend-ID}.
Each behavior record contains \emph{user-ID}, \emph{venue-ID}, \emph{timestamp} and \emph{UGC}.
Their basic statistics are shown in Table \ref{t1}.
%, as illustrated in Table \ref{example-foursquare}.
%%%%%%%%%%%%%%%%%%%%%%%%%%%%%%%%%%%%%%%%%%%%%%%%%期刊版恢复%%%%%%%%%%%%%%%%%%%%%%%%%%%%%%%%%%%%%%%%%%%%%%%%%%%%%%%%%%%%%%%%%
%\begin{table*}[t]\renewcommand{\arraystretch}{1.3}
%\centering
%\caption{Part of A User's Behavior Records in Foursquare Dataset}\label{example-foursquare}
%\scalebox{0.9}{\begin{tabular}{p{1.5cm}| p{3.8cm} |p{1.8cm} | p{9.3cm}}
%% \diagbox{Height}{Group}{BMI}&\(<18\)&\(18\sim22\)&\(>22\)\\
%\hline
%\textbf{User-ID}&\textbf{Venue-ID}&\textbf{Timestamp}&\textbf{Message Content}  \\
% \hline
%  \hline
%\ 2287345 &4d029c1254d0236a94e2e3d5& 1299135219 &Pen is better!  The class sizes are only 20:1!!!  GO PANTHERS!\\
%\ 2287345 &4c7e886601df37046186e3ac& 1299135270 &GO PANTHERS!\\
%\ 2287345 &4c9a2eba80958cfa40a940d4& 1299135328 &Save a whale, eat a Sea King!\\
%\ 2287345 &4c7c6b4fdbaa76b04e1c314b& 1301004689 &Best school in the world.  PV High is an uber fail compared to here.\\
%\ 2287345 &4c7c6b4fdbaa76b04e1c314b& 1303711907 &The best teachers only come from Pen.\\
%\ 2287345 &4db7abcc6a238ed5d8e486d8& 1303971421 &DJ Mike too the rescue!\\
%%\ \# of social ties&330,898&1,262,659\\
%\hline
%\end{tabular}
%}
%\end{table*}
%%%%%%%%%%%%%%%%%%%%%%%%%%%%%%%%%%%%%%%%%%%%%%%%%期刊版恢复%%%%%%%%%%%%%%%%%%%%%%%%%%%%%%%%%%%%%%%%%%%%%%%%%%%%%%%%%%%%%%%%%

\begin{figure}
\begin{center}
\begin{tabular}{cc}
  %\scalebox{0.5}{\includegraphics[width=0.45 \textwidth]{Foursquare-user-recordnumber-distribution.eps}}&
%  \scalebox{0.5}{\includegraphics[width=0.45 \textwidth]{Yelp-user-recordnumber-distribution.eps}}\\
  \scalebox{0.5}{\includegraphics[width=0.45 \textwidth]{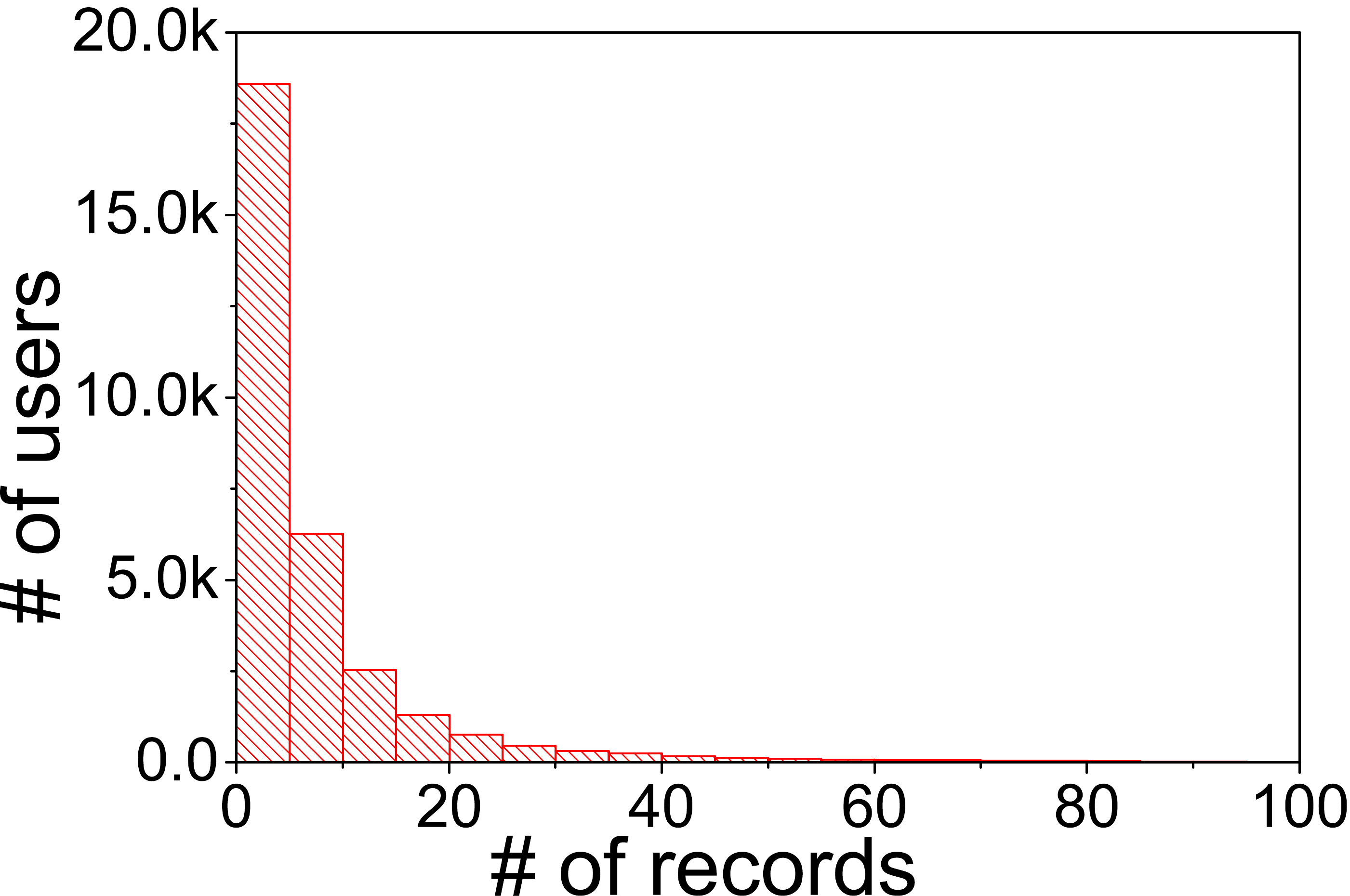}}&
  \scalebox{0.5}{\includegraphics[width=0.45 \textwidth]{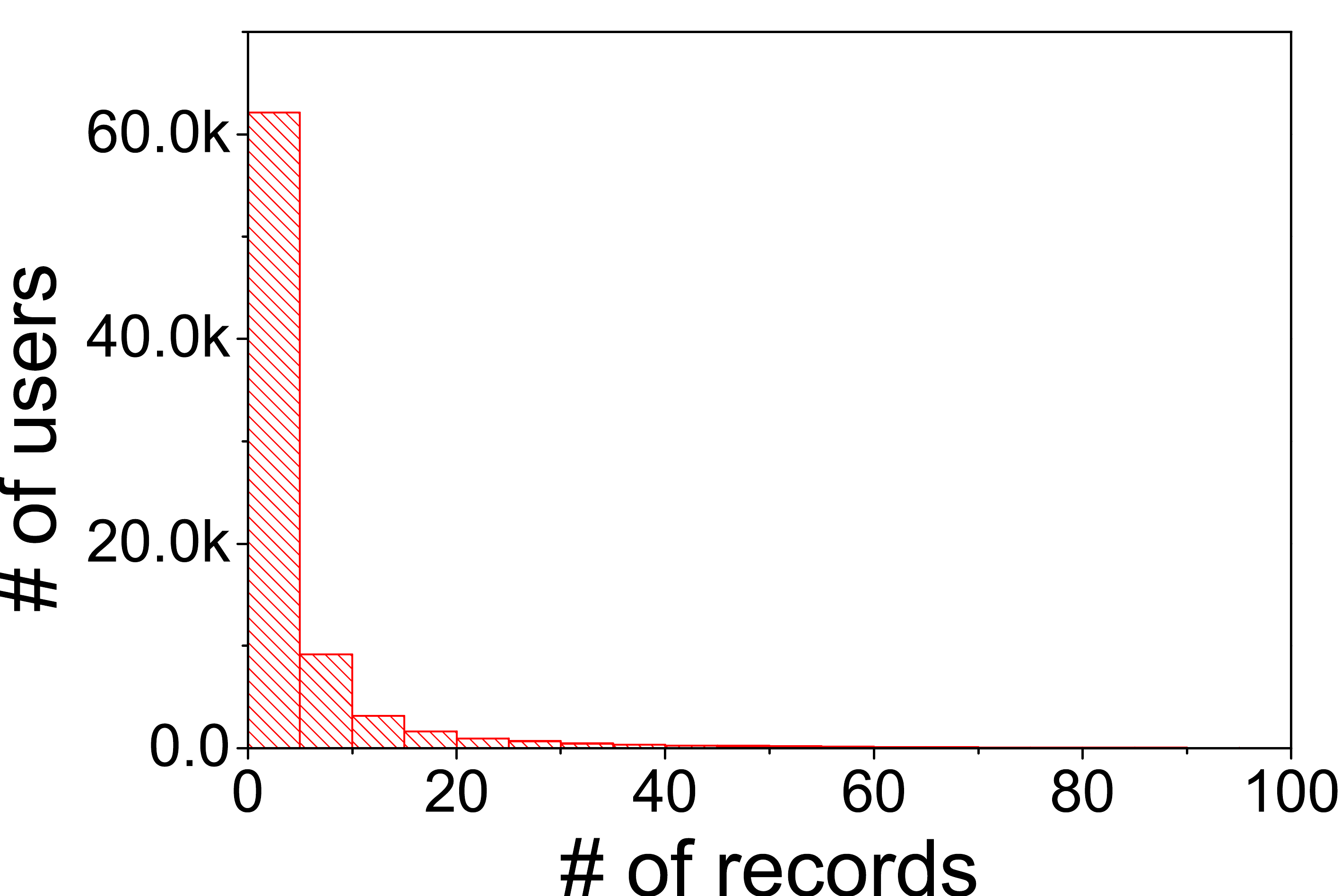}}\\
 {\small(a) Foursquare.} & {\small (b) Yelp.}
\end{tabular}
\end{center}
\vspace{-0.1in}
\caption{The distribution of user record counts.}\label{F-user-record-distribution}
%\vspace{-0.1in}
\end{figure}

We count each user's records, and present the results in Fig. \ref{F-user-record-distribution}.
It shows that most users have less than $5$ records in both datasets.
The quality of these dataset is too \emph{poor} to model individual-level behavioral patterns for the majority of users, which confronts our method with a big challenge.

%But our experimental results indicate that our joint model has a good performance in this hard scenario.

%\begin{figure}
%\centering
 % \includegraphics[width=0.45 \textwidth]{Yelp-user-recordnumber-distribution.eps}
 % \caption{The distribution of user record counts in Yelp dataset.}\label{Y-user-record-distribution}
%\end{figure}

%\vspace{-0.05in}
\subsection{Experiment Settings}\label{Experiment-analysis}
\subsubsection{Suspicious Behavior Simulation}\label{subsubsection-susuipdocuou-buhaofbjou}

Many works \cite{EgeleSKV17, ViswanathBCGGKM14, LesaegeSLV16} aimed at discovery theft's behavioral pattern.
Bursztein et al. \cite{BurszteinBMPAAPS14} pointed out that identity thieves usually behave in two possible suspicious patterns, i.e.,
%(1) ones behave unlike the majority of users;
%(2) ones behave only unlike the victim.
(1) behaving unlike the majority of users;
(2) behaving only unlike the victim.
Many existing outlier detecting techniques, e.g., i-Forest \cite{LiuTZ08}, LOF \cite{YanCKR17} and GSDPMM \cite{YinW16} can deal with the former cases.
Besides, we notice that the former can be regarded as a special case of the latter.
It is straightforward that an effective detection method for the latter can apply effectively to the former cases.
If the experiments validate that our model performs well even for detecting such crafty thieves, a strong argument can be obtained to prove the capability of our model.
This is the reason why we focus on the latter cases where thieves tend to hind them among the people.

In the experiments, we use two real-life datasets, and assume that all records are normal behaviors.
We simulate suspicious behaviors by exchanging some users' behavioral records and setting them as positive instances \cite{BurszteinBMPAAPS14}.
This theft simulation process imitates one kind of the most crafty thieves who behave just like normal users.
More specifically, we first rank  behavior records according to their timestamps.
Then, we select the top $80\%$  behavior records for training and the rest for testing.
To simulate suspicious behaviors, we randomly exchange $5\%$ of all behavior records in the test set as \emph{anomalous behaviors}.
Totally, we have $56,236$ test behaviors in Foursquare and $71,667$ in Yelp, and make up $2,884$ anomalous behaviors in Foursquare and $3,583$ in Yelp, respectively.

\begin{figure}
\begin{center}
\begin{tabular}{cc}
% \scalebox{0.5}{\includegraphics[width=0.45 \textwidth]{Foursquare-joint-AS-normal-compromised-histogram.eps}}&
% \scalebox{0.5}{\includegraphics[width=0.45 \textwidth]{Yelp-joint-AS-normal-compromised-histogram.eps}}
\scalebox{0.5}{\includegraphics[width=0.45 \textwidth]{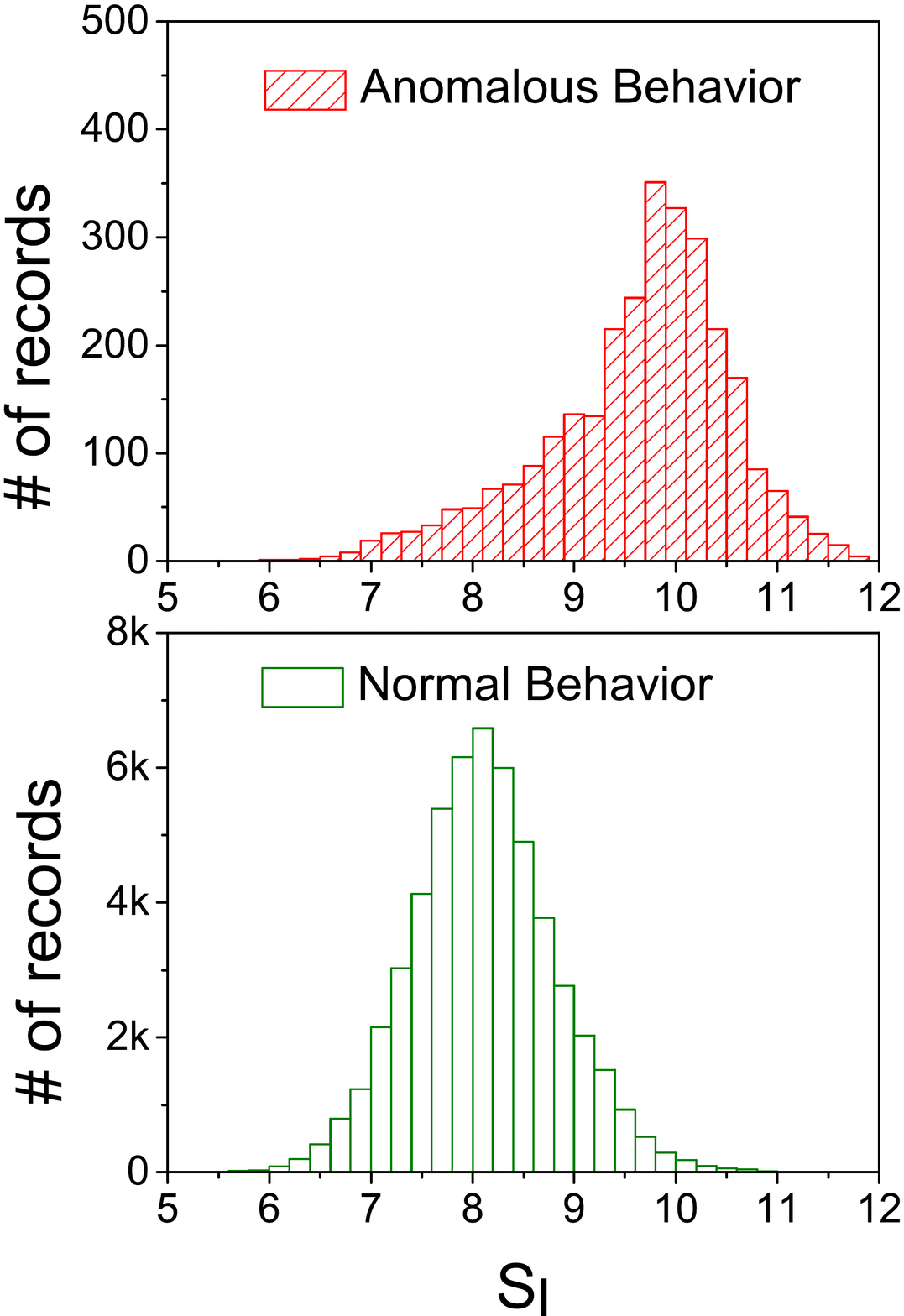}}&
 \scalebox{0.5}{\includegraphics[width=0.45 \textwidth]{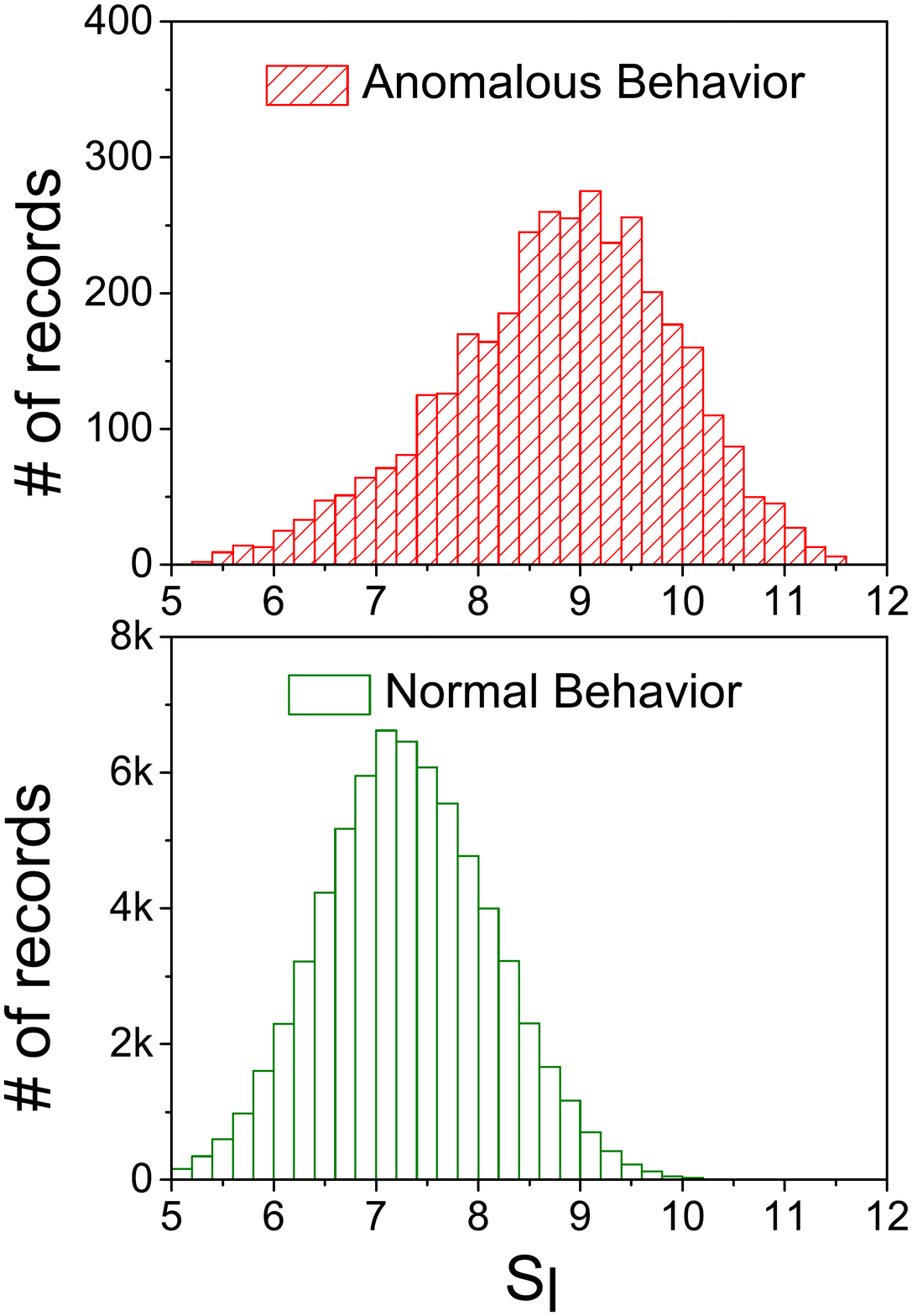}}
\\
{\small(a) Foursquare} & {\small (b) Yelp}
\end{tabular}
\end{center}
\vspace{-0.1in}
\caption{The histogram of logarithmic anomalous score $S_l$ (defined in Eq. (\ref{BR})) for each behavior.}\label{F-AS}
%\vspace{-0.2in}
\end{figure}

\begin{figure}
\begin{center}
\begin{tabular}{cc}
 % \scalebox{0.5}{\includegraphics[width=0.45 \textwidth]{Foursquare-joint-ras-normal-compromised-histogram.eps}}&
%  \scalebox{0.5}{\includegraphics[width=0.45 \textwidth]{Yelp-joint-ras-normal-compromised-histogram.eps}}
  \scalebox{0.5}{\includegraphics[width=0.45 \textwidth]{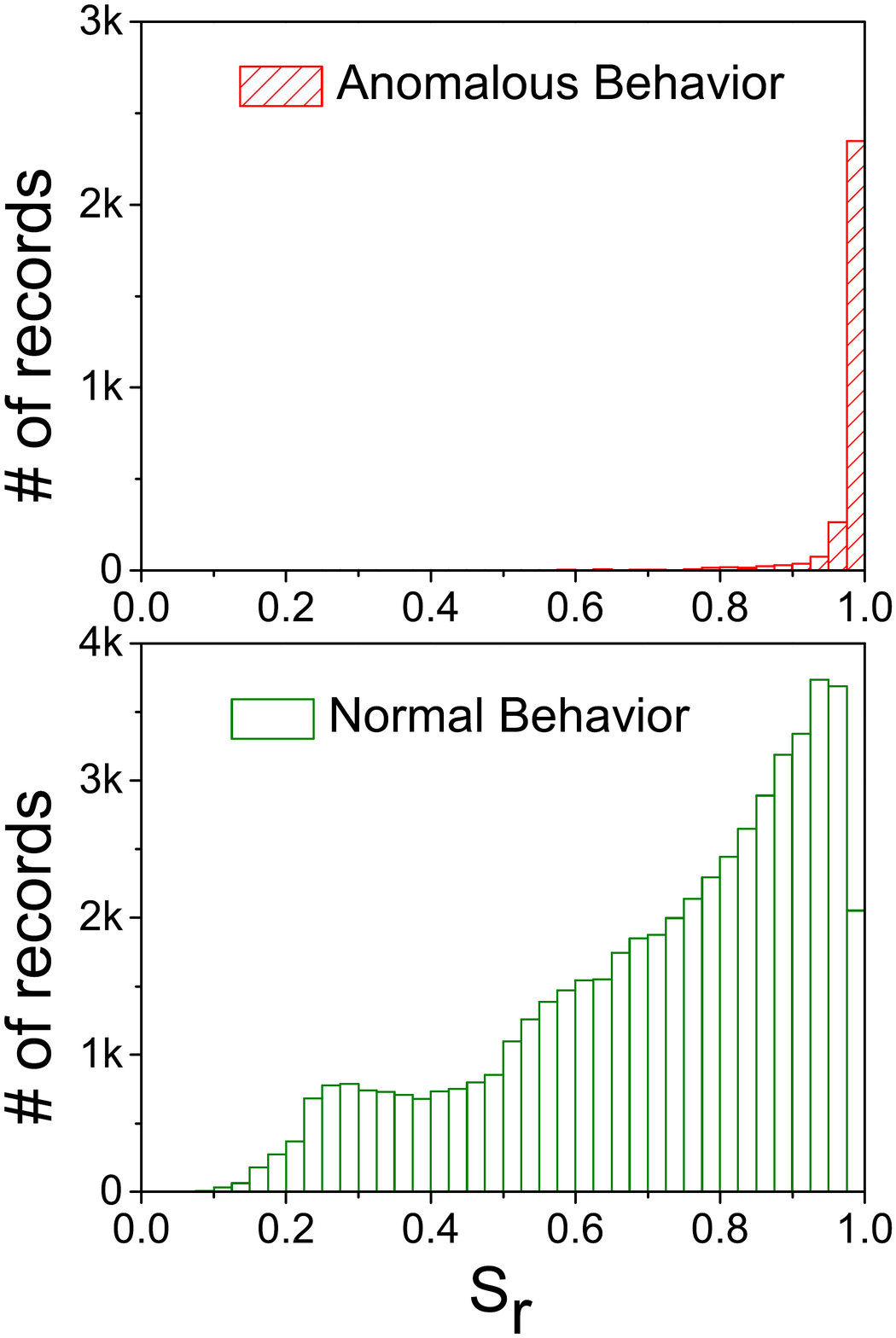}}&
  \scalebox{0.5}{\includegraphics[width=0.45 \textwidth]{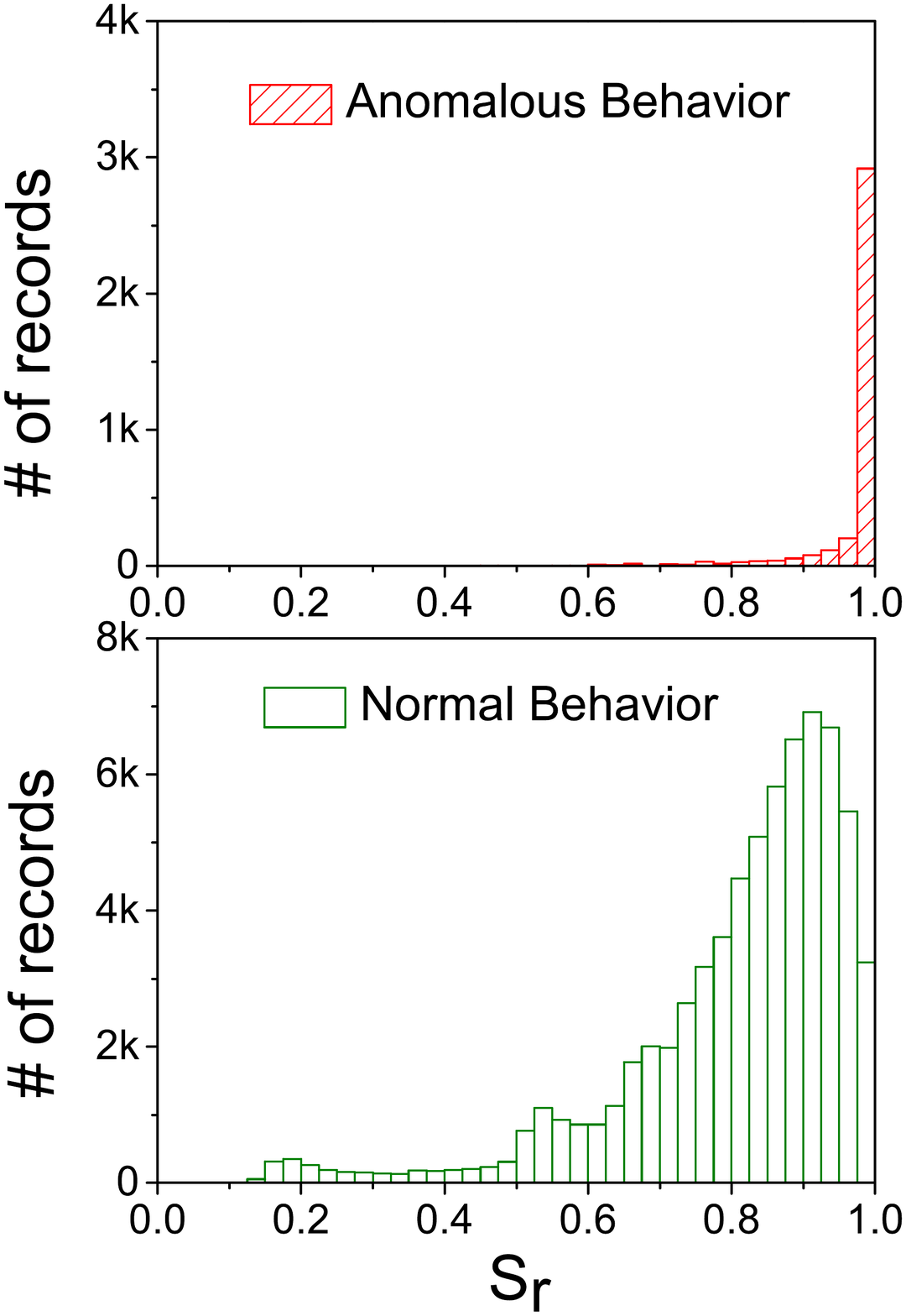}}
\\
{\small(a) Foursquare} & {\small (b) Yelp}
\end{tabular}
\end{center}
%\vspace{-0.15in}
  \caption{The histogram of relative anomalous score $S_r$ (defined in Eq. (\ref{RBR})) for each behavior.}\label{F-RAS}
  %\vspace{-0.15in}
\end{figure}

\subsubsection{Metrics}

For the convenience of description, we first give a confusion matrix in Table \ref{confusion-matrix}.
\begin{table}[h]\renewcommand{\arraystretch}{1.2}
\caption{Confusion Matrix for Binary Classification.}\label{confusion-matrix}
\vspace{-0.1in}
\scalebox{0.9}{\begin{tabular}{p{2.6cm}|p{2.8cm}|p{2.8cm}}
\toprule
\multirow{2}{*}{Predicted Condition}&\multicolumn{2}{c}{True Condition}\cr\cline{2-3}
&Positive&Negative\cr
\hline
Positive&True Positive (TP)&False Positive (FP)\\
\hline
Negative&False Negative (FN)&True Negative (TN)\\
\bottomrule
\end{tabular}
}
%\vspace{-0.1in}
\end{table}

In the experiments, we set \emph{anomalous behaviors} as \emph{positive instances}, and focus on the following four metrics, since the identity theft detection is essentially an \emph{imbalanced binary classification problem} \cite{HeG09}.

\vspace{+0.02in}
\textbf{True Positive Rate} (TPR): TPR is computed by $\frac{TP}{TP+FN}$, and indicates the proportion of true positive instances in all positive instances (i.e.,  the proportion of anomalous behaviors that are detected in all anomalous behaviors).
It is also known as \emph{recall}.
Specifically, we named it \emph{detection rate}.
%\begin{equation}\label{TPR}
%TPR=\frac{TP}{TP+FN}
%\end{equation}

\textbf{False Positive Rate} (FPR): FPR is computed by $\frac{FP}{FP+TN}$, and indicates the proportion of false positive instances in all negative instances (i.e.,  the proportion of  normal behaviors that are mistaken for anomalous behaviors in all normal behaviors).
Specifically, we named it \emph{disturbance rate}.
%\begin{equation}\label{FPR}
%TPR=\frac{FP}{FP+TN}
%\end{equation}

\textbf{Precision}: The precision is computed by $\frac{TP}{TP+FP}$, and indicates the proportion of true positive instance in all predicted positive instance (i.e.,  the proportion of  anomalous behaviors that are detected in all suspected cases).
%\begin{equation}\label{Precision}
%P=\frac{TP}{TP+FP}
%\end{equation}

\textbf{AUC}: Given a rank of all test behaviors, the AUC value can be interpreted as the probability that a classifier/predictor will rank a randomly chosen positive instance higher than a randomly chosen negative one.

%\vspace{-0.05in}
\subsubsection{Threshold Selection}
It is an important issue in classification tasks.
Recall that for the hyper-parameters $\alpha$, $\beta$, $\gamma$ and $\eta$, we adopt a fixed value, i.e., $\alpha = 50/Z$, $\gamma = 50/C$ and $\beta = \eta = 0.01$, following the study \cite{YinHZWZHS16}.
Specifically, we take the case that $C=30$ and $Z=20$ as an example to present the threshold selection strategy.
The parameter sensitivity analysis will be conducted  in the following Section \ref{subsub-paerjmentsenttivanal}.
%Here, we use the experimental result on the Foursquare dataset as an example to introduce our threshold selection strategy.
We compare the distribution of \emph{logarithmic anomalous score} $S_l$ (or \emph{relative anomalous score} $S_r$) for normal behaviors with that for anomalous behaviors.
Figs. \ref{F-AS} and \ref{F-RAS} present the differences between normal and anomalous behaviors in terms of the distributions of $S_l$ and $S_r$, respectively.
They show that the differences are both significant, and the difference in terms of $S_r$ is much more obvious.

\begin{figure}
\begin{center}
\begin{tabular}{cc}
 %\scalebox{0.5}{\includegraphics[width=0.45 \textwidth]{Foursquare-joint-ras-limit-normal-compromised-histogram.eps}}&
% \scalebox{0.5}{\includegraphics[width=0.45 \textwidth]{Yelp-joint-ras-limit-normal-compromised-histogram.eps}}
 \scalebox{0.5}{\includegraphics[width=0.45 \textwidth]{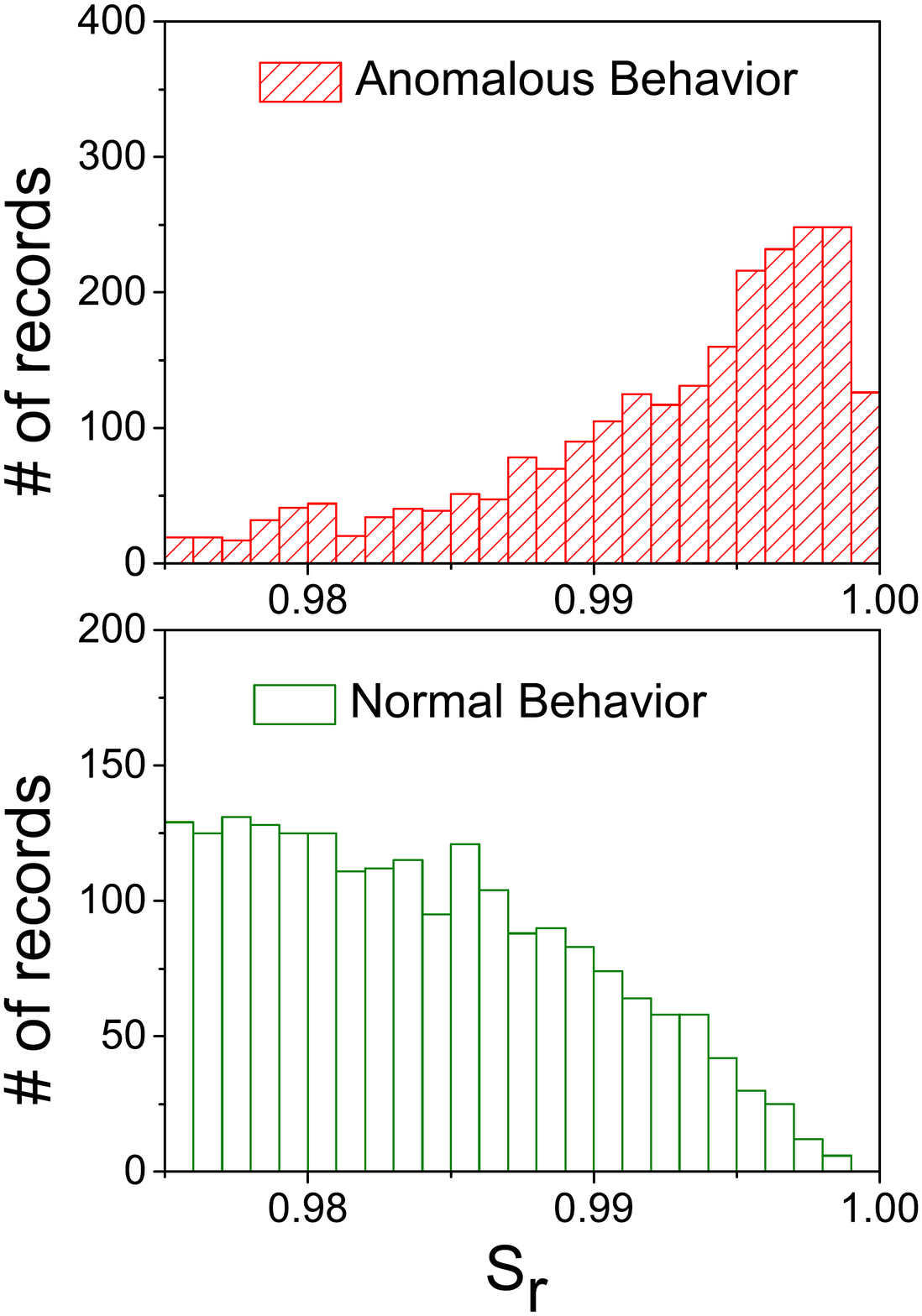}}&
 \scalebox{0.5}{\includegraphics[width=0.45 \textwidth]{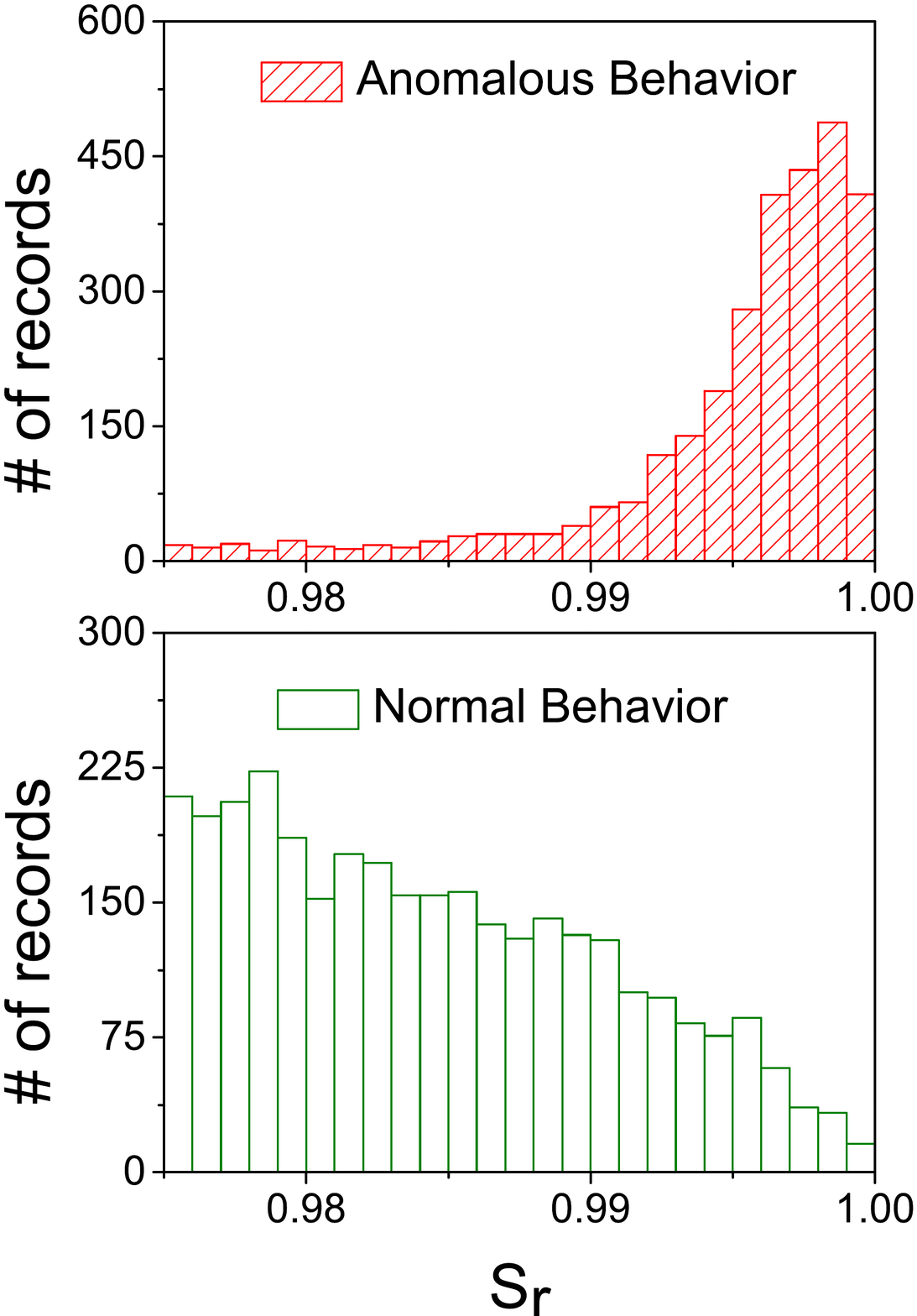}}
\\
{\small(a) Foursquare} & {\small (b) Yelp}
\end{tabular}
\end{center}
%  \vspace{-0.15in}
  \caption{A partial of the distribution of relative anomalous score $S_r$ (defined in Eq. (\ref{RBR})) for each behavior.}\label{F-RAS-part}
%\vspace{-0.15in}
\end{figure}

\begin{figure}
\begin{center}
\begin{tabular}{cc}
  %\scalebox{0.5}{\includegraphics[width=0.45 \textwidth]{Foursquare-joint-ras-limit-ROC.eps}}&
%  \scalebox{0.5}{\includegraphics[width=0.45 \textwidth]{Yelp-joint-ras-limit-ROC.eps}}
  \scalebox{0.5}{\includegraphics[width=0.45 \textwidth]{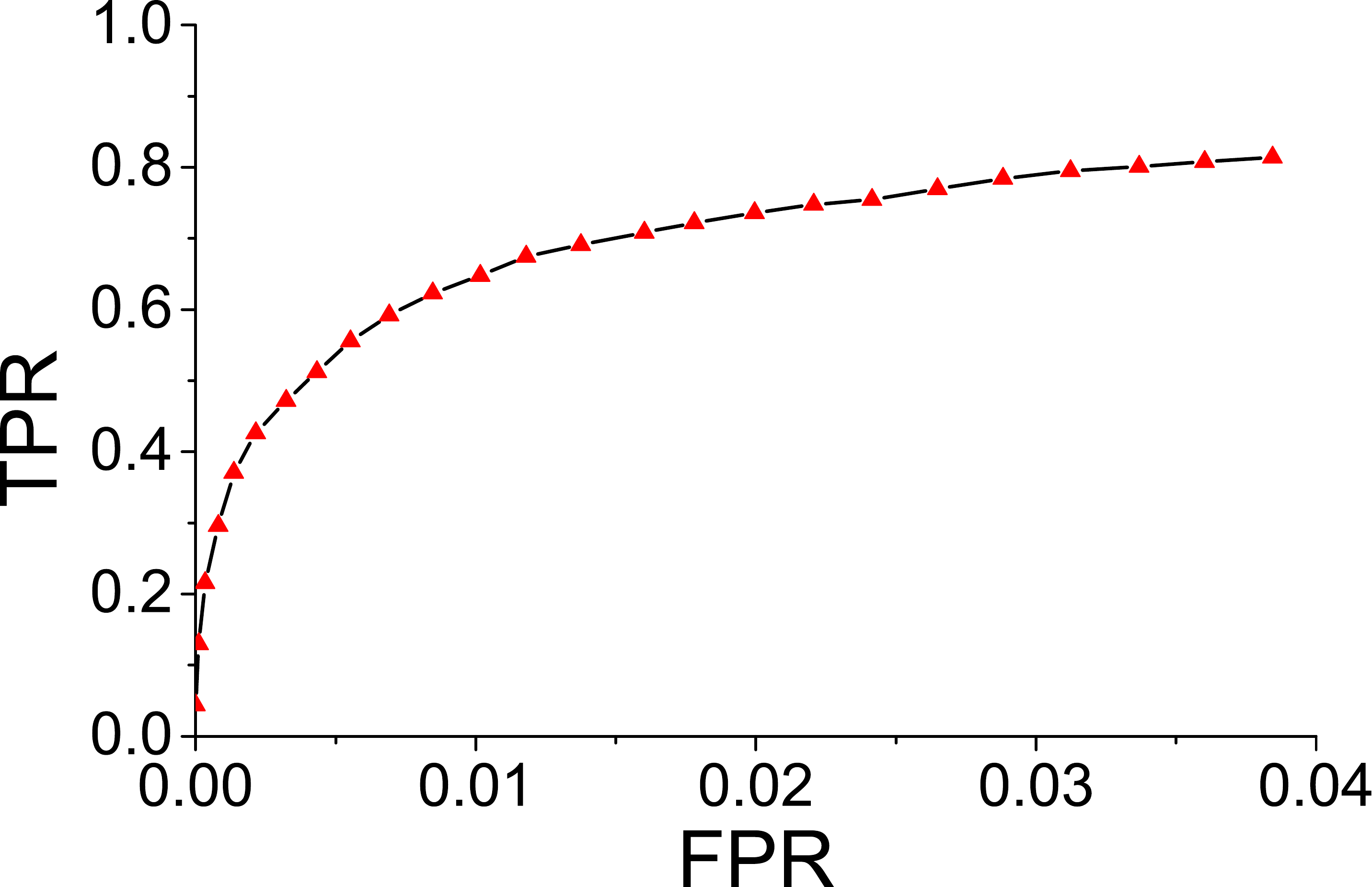}}&
  \scalebox{0.5}{\includegraphics[width=0.45 \textwidth]{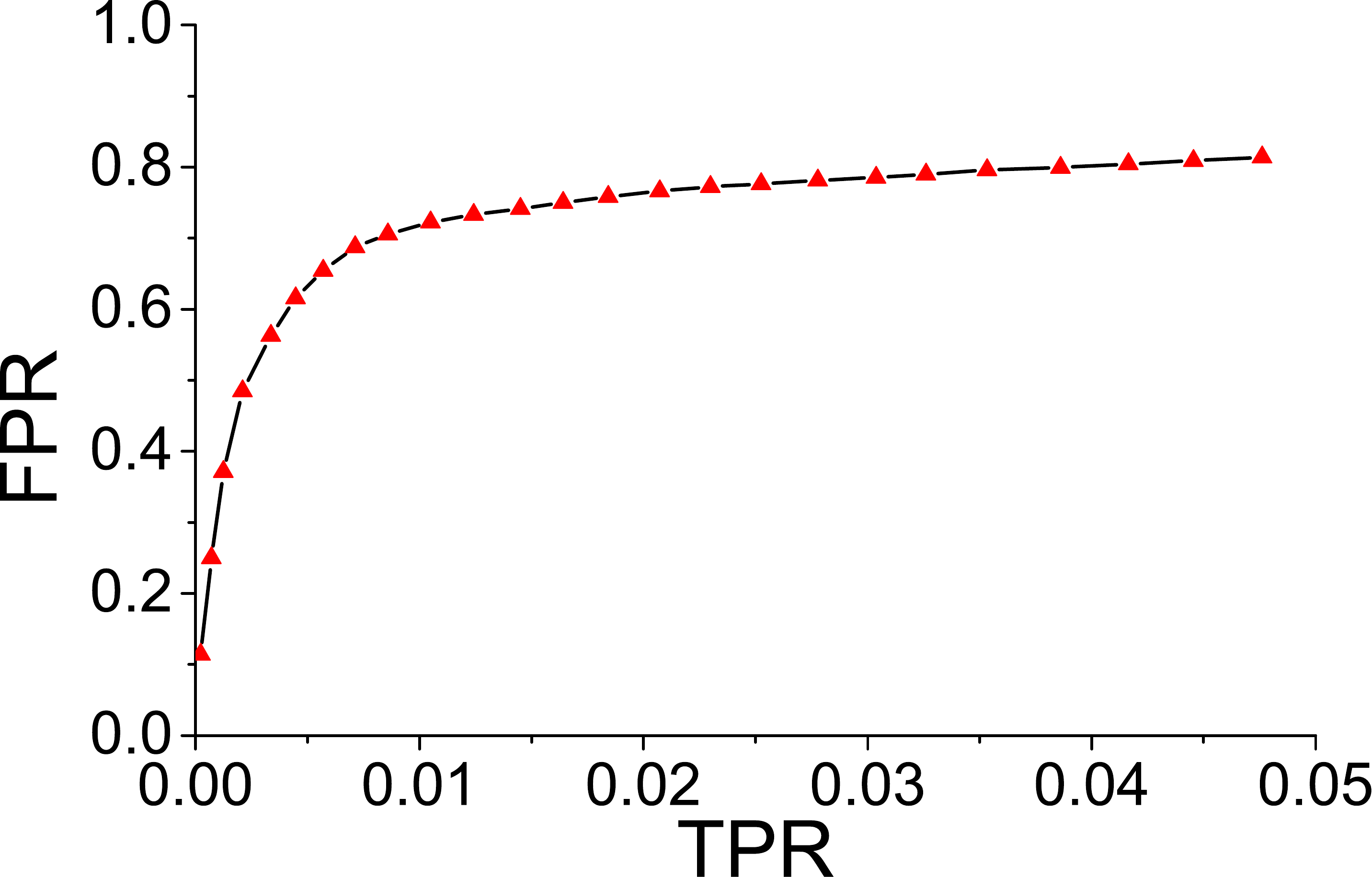}}
\\
{\small(a) Foursquare} & {\small (b) Yelp}
\end{tabular}
\end{center}
% \vspace{-0.15in}
  \caption{A partial of  ROC (receiver operating characteristic) curve of identity theft detection.}\label{F-ROC-part}
  % \vspace{-0.05in}
\end{figure}

\begin{figure}[t]
%\begin{center}
\begin{tabular}{cc}
  %\scalebox{0.5}{\includegraphics[width=0.45 \textwidth]{Foursquare-joint-threshold-cost.eps}}&
%  \scalebox{0.5}{\includegraphics[width=0.45 \textwidth]{Yelp-joint-threshold-cost.eps}}
  \scalebox{0.5}{\includegraphics[width=0.45 \textwidth]{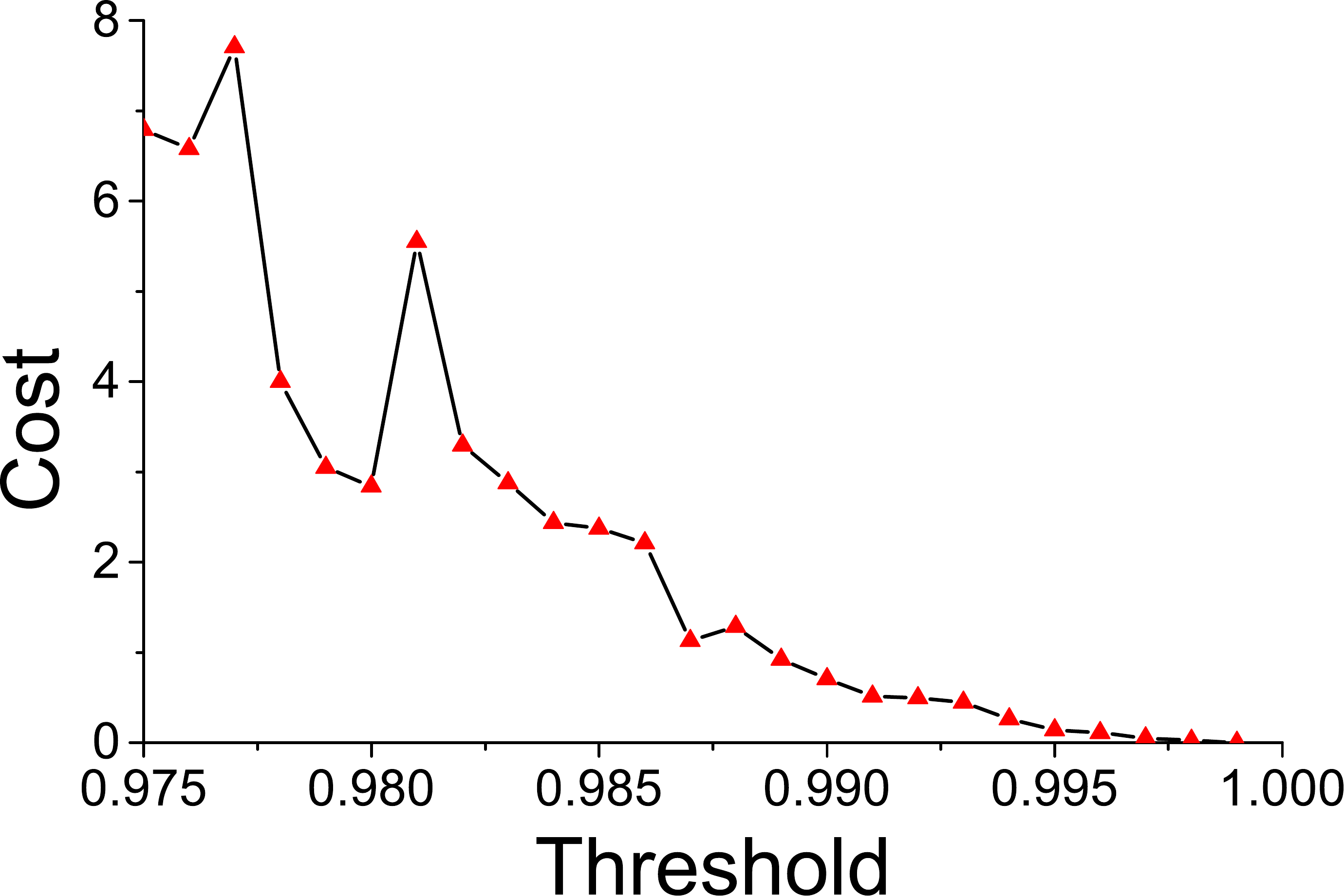}}&
  \scalebox{0.5}{\includegraphics[width=0.45 \textwidth]{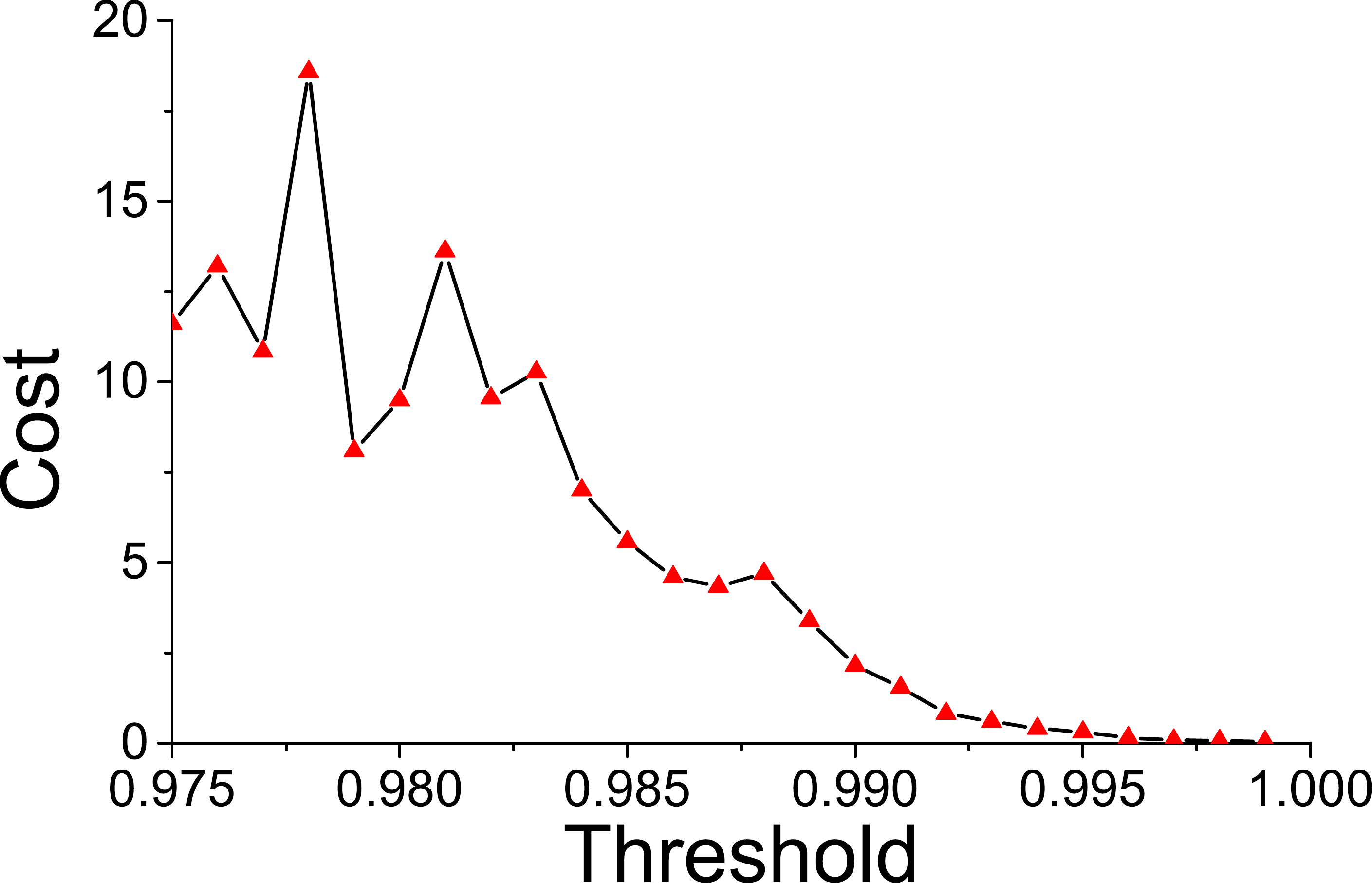}}
\\
{\small(a) Foursquare} & {\small (b) Yelp}
\end{tabular}
%\end{center}
 %\vspace{-0.1in}
  \caption{Detection costs with different thresholds.}\label{F-cost}
  % \vspace{-0.1in}
\end{figure}

To obtain a reasonable threshold, we focus on the performance where the threshold changes from $0.975$ to $1$, since this range contains $81.5\%$ ($81.4\%$) of all anomalous behaviors and $3.9\%$ ($4.8\%$) of all normal behaviors in Foursquare (Yelp).
The detailed trade-offs are demonstrated in Figs. \ref{F-RAS-part} and \ref{F-ROC-part} from different aspects.
To optimize the trade-offs of detection performance,
we define the detection \emph{Cost} in Eq. (\ref{cost}):
\begin{equation}
Cost=\frac{\# \mbox{ of newly mistaken normal behaviors}}{\# \mbox{ of newly identified anomalous behaviors}}. \label{cost}
\end{equation}
We present the threshold-cost curve in Fig. \ref{F-cost}.
It shows that a smaller threshold usually corresponds to a larger cost.

We select the minimum threshold satisfying that the corresponding cost is less than $1$.
Thus, we choose $0.989$ and $0.992$ as the thresholds for Foursquare and Yelp, respectively.
Under them, our joint model CBM reaches $62.32\%$ ($68.75\%$) in TPR and $0.85\%$ ($0.71\%$) in FPR on Foursquare (Yelp).
Please refer to Table \ref{metrics-summary} for details.

\begin{table}[t]\renewcommand{\arraystretch}{1.3}
\centering
\caption{A Summary of Different Metrics \cite{Hastie2008} with the Threshold $0.989$ for Foursquare and $0.992$ for Yelp.}\label{metrics-summary}
%\vspace{-0.1in}
\begin{tabular}{p{2cm}|p{2.6cm}p{2.6cm}}
\toprule
&Foursquare&Yelp\\
\hline
 \ \textbf{AUC}&0.956&0.947\\
 \ \textbf{Precision}&79.91\%&83.55\%\\
 \ \textbf{Recall (TPR)}&62.32\%&68.75\%\\
 \ \textbf{FPR}&0.85\%&0.71\%\\
 \hline
 \ TNR&99.15\%&99.29\%\\
 \ FNR&37.68\%&31.25\%\\
 \ Accuracy&97.26\%&97.76\%\\
 \ F1&0.700&0.754\\
\bottomrule
\end{tabular}
\vspace{-0.1in}
\end{table}

%\vspace{-0.05in}
\subsubsection{Parameter Sensitivity Analysis}\label{subsub-paerjmentsenttivanal}
Parameter tuning is another important part of our work.
The performance of our model is indeed sensitive to the number of communities ($C$) and topics ($Z$).
Therefore, we study the impact of varying parameters in our model.
We select the relative anomalous score $S_r$ as the test variable, and evaluate the performance of our model by changing the values of $C$ and $Z$.
The experimental results are summarized in Tables \ref{CZ-F}.
\begin{table}[t]\renewcommand{\arraystretch}{1.3}
\centering
\caption{AUC on Foursquare (Yelp) Dataset}\label{CZ-F}
%\vspace{-0.1in}
\begin{tabular}{ p{1.5cm} | p{1.8cm} p{1.8cm} p{1.8cm}}
\toprule
&C=10&C=20&C=30\\
\hline
\ Z=10&0.876 (0.910) &0.945 (0.936) &0.953 (0.945) \\
\ Z=20&0.917 (0.915) &0.946 (0.938) &\textbf{0.956} (\textbf{0.947})\\
\ Z=30&0.922 (0.917) &0.947 (0.938) &0.957 (0.947) \\
\bottomrule
\end{tabular}
%\vspace{-0.15in}
\end{table}

\begin{figure}[t]
\begin{center}
\begin{tabular}{cc}
\scalebox{0.5}  {\includegraphics[width=0.45 \textwidth]{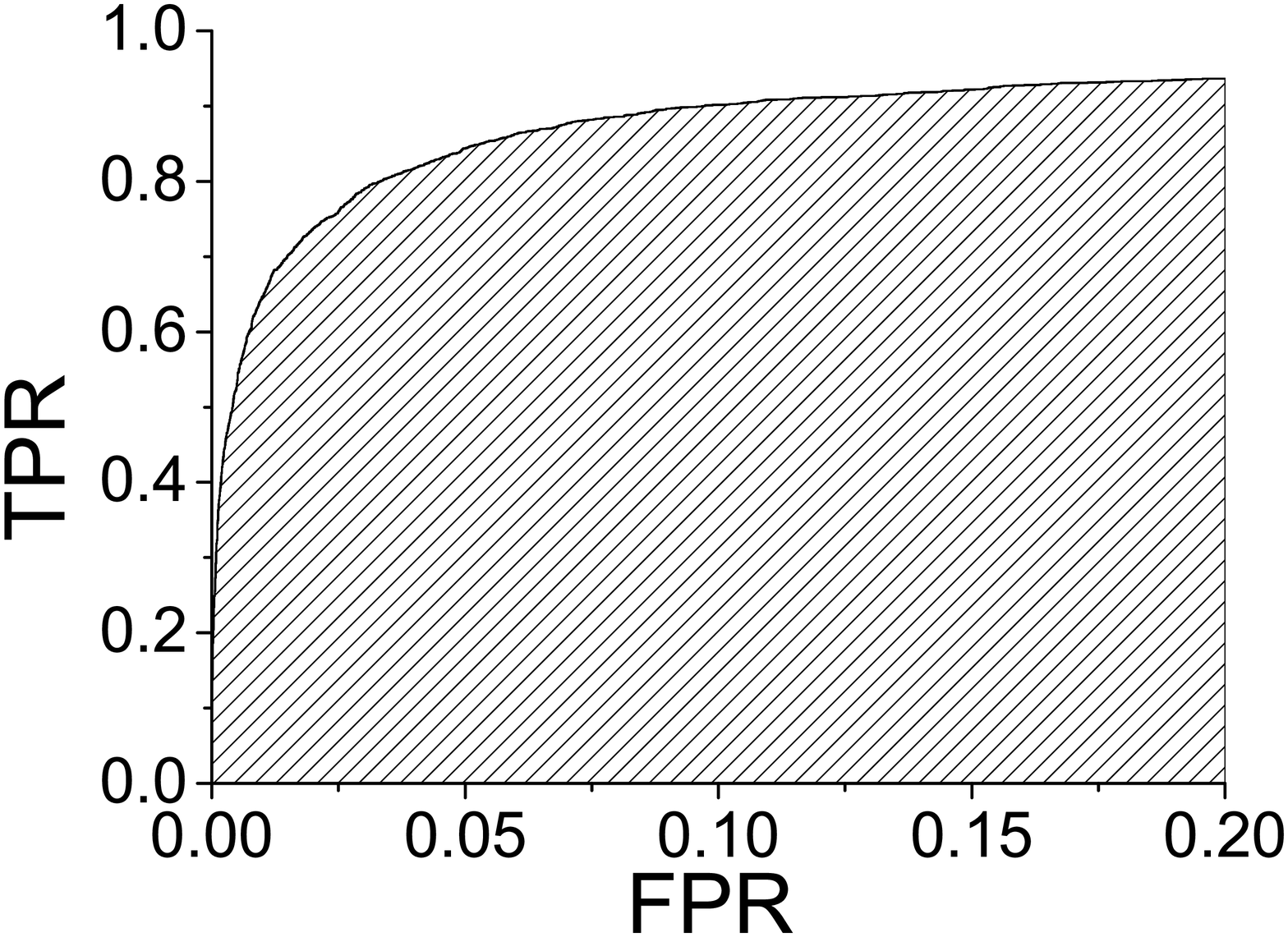}}&
% % \caption{The ROC curve of identity theft detection via joint model on Foursquare dataset.}\label{F-j1}
 \scalebox{0.5} {\includegraphics[width=0.45 \textwidth]{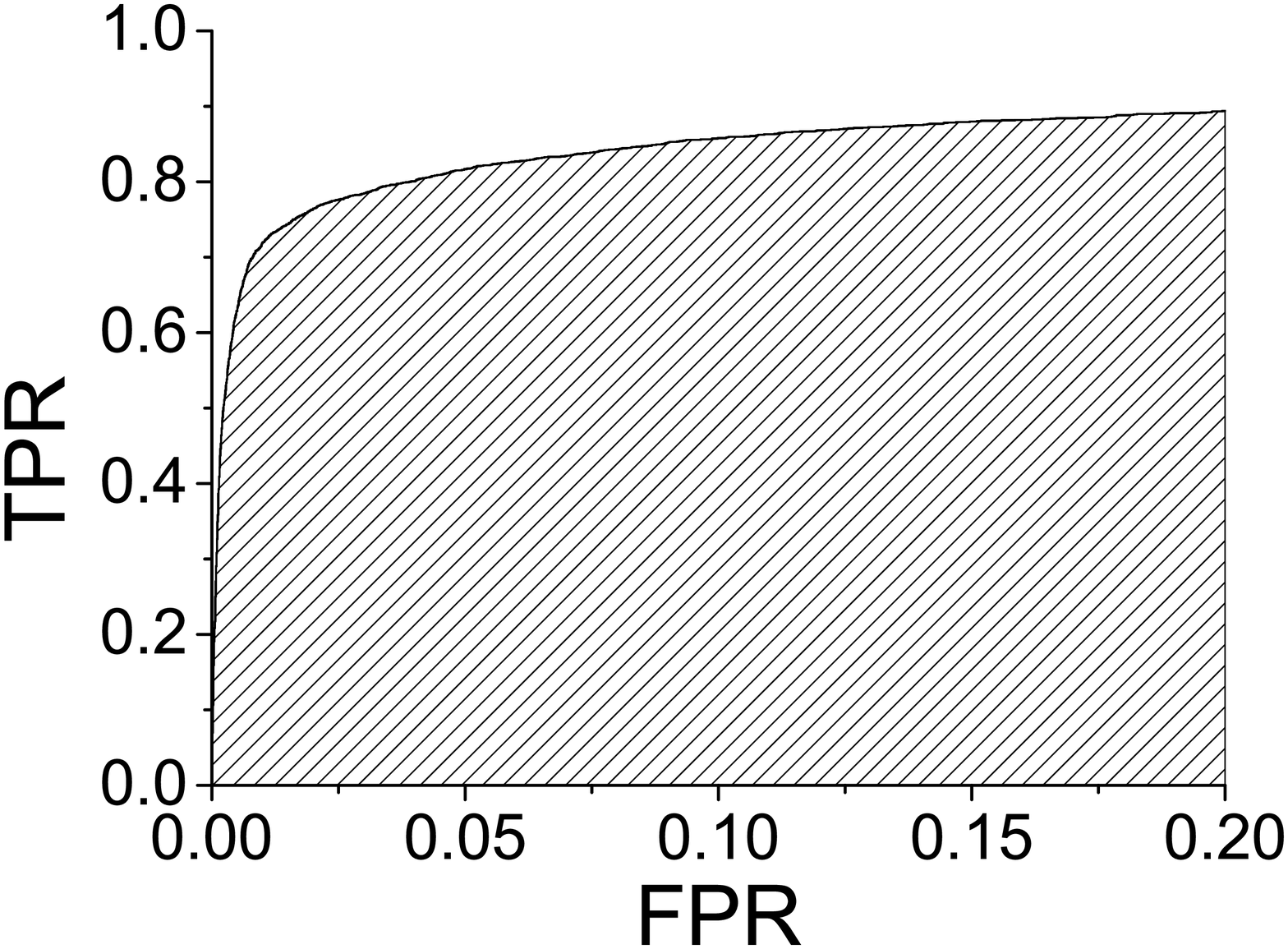}}\\
% \scalebox{0.5}  {\includegraphics[width=0.45 \textwidth]{Foursquare-joint-c30z20-1-ROC.pdf}}&
 % \caption{The ROC curve of identity theft detection via joint model on Foursquare dataset.}\label{F-j1}
% \scalebox{0.5} {\includegraphics[width=0.45 \textwidth]{Yelp-joint-c30z20-1-ROC.pdf}}\\
 {\small(a) Foursquare} & {\small (b) Yelp}
%  \caption{The ROC curve of identity theft detection via joint model on Yelp dataset.}\label{Y-j1}
\end{tabular}
\end{center}
\vspace{-0.1in}
\caption{The ROC curves of identity theft detection via the joint model CBM.} \label{ROC-curve}
%\vspace{-0.1in}
\end{figure}

\begin{figure}[t]
\begin{center}
\begin{tabular}{cc}
%\scalebox{0.5}{\includegraphics[width=0.45 \textwidth]{Foursquare-joint-c30z20-1-PR.eps}} & \scalebox{0.5}{\includegraphics[width=0.45 \textwidth]{Yelp-joint-c30z20-1-PR.eps}}\\
\scalebox{0.5}{\includegraphics[width=0.45 \textwidth]{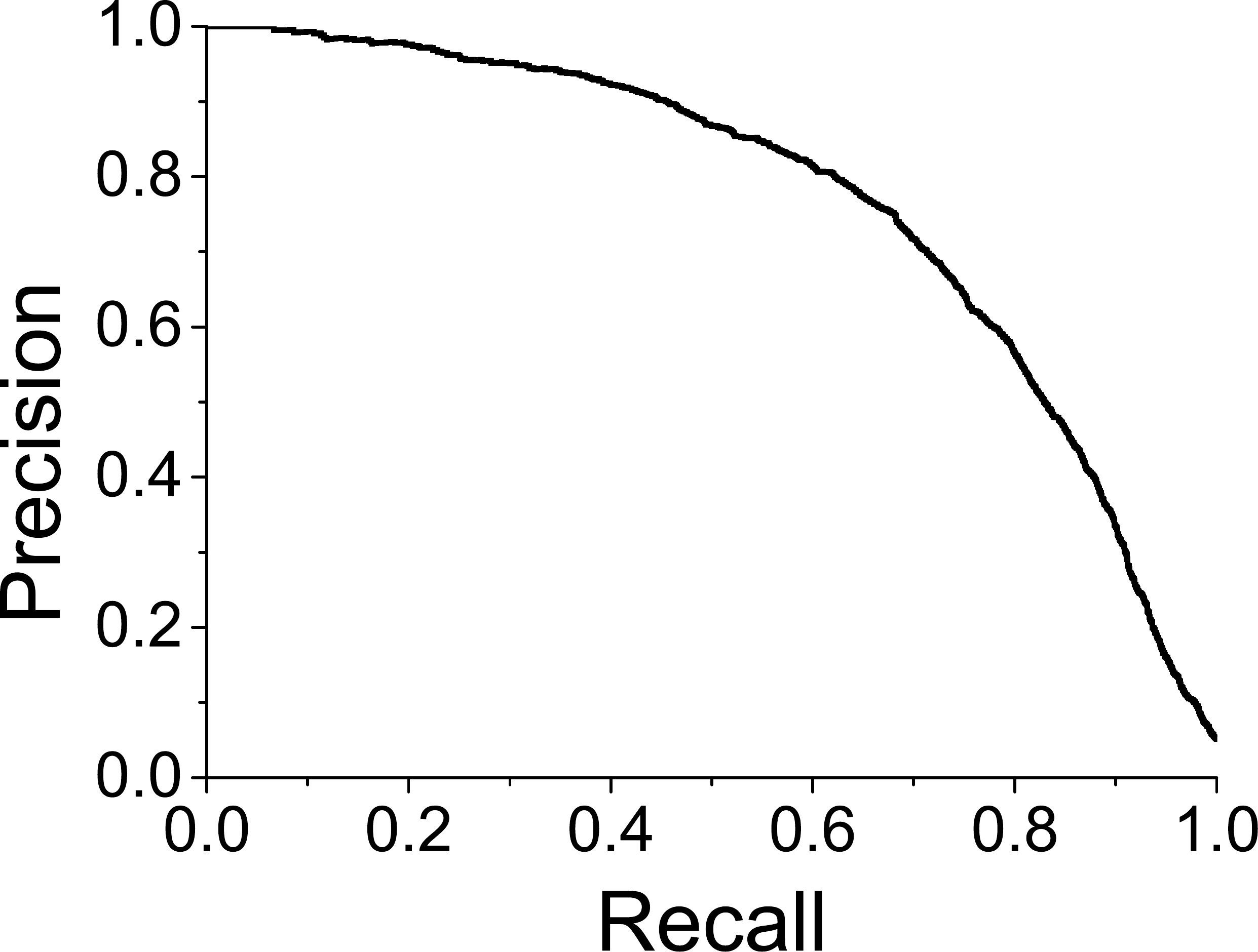}} & \scalebox{0.5}{\includegraphics[width=0.45 \textwidth]{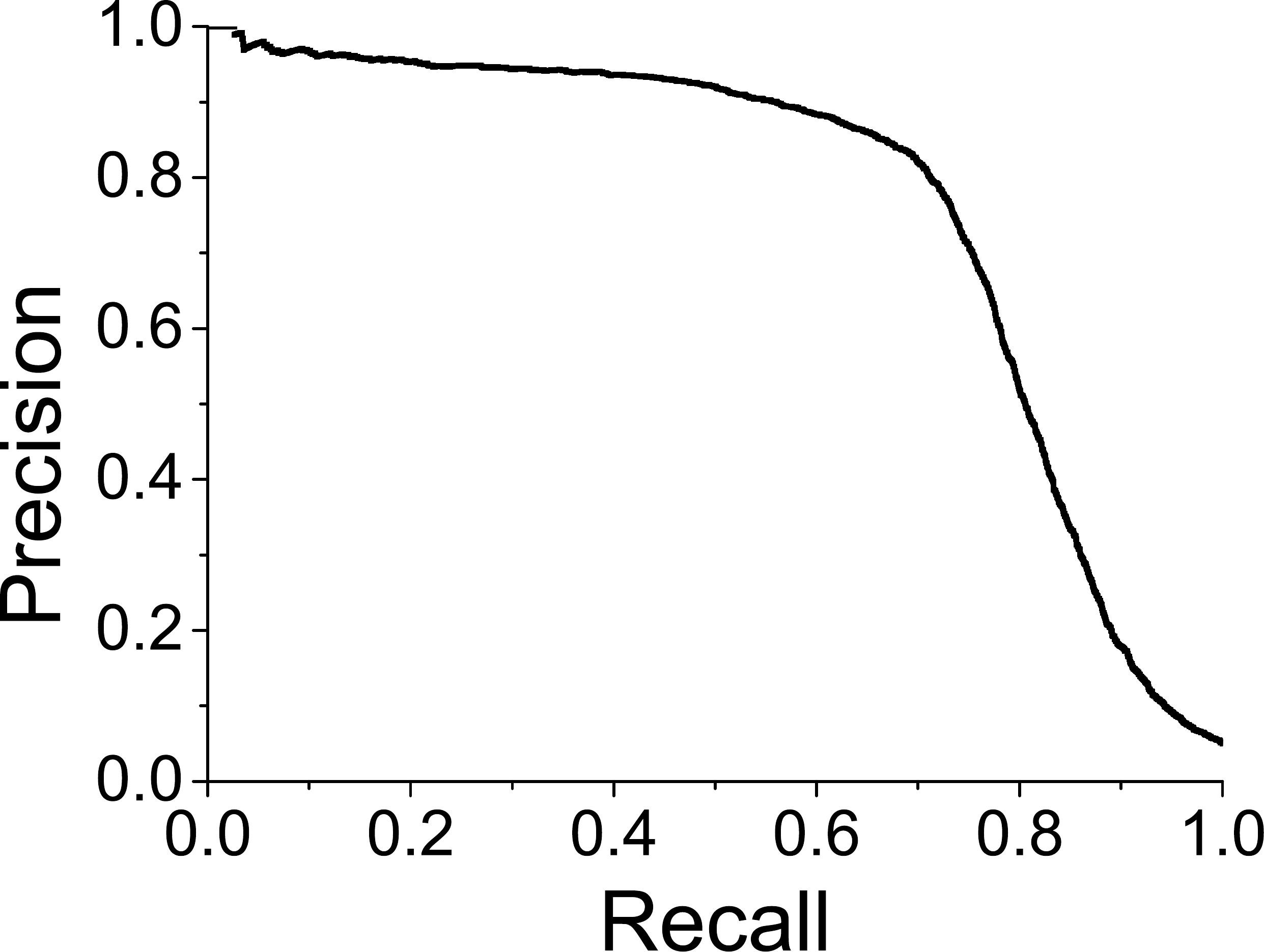}}\\
{\small(a) Foursquare} & {\small (b) Yelp}
\end{tabular}
\end{center}
\vspace{-0.1in}
\caption{The Precision-Recall curves of identity theft detection via the joint model CBM.} \label{PR-curve}
%\vspace{-0.05in}
\end{figure}

From the results on both datasets, the detection efficacy goes stable when the number of topics reaches $20$ and the number of communities has a larger impact on the efficacy.
Thus, we set $C=30$ and $Z=20$ in our joint model, and present the receiver operating characteristic (ROC) and Precision-Recall curves in Figs. \ref{ROC-curve} and \ref{PR-curve}, respectively.
Specifically, we present detection rate (TPR) in Table \ref{Efficacy}, where disturbance rate (FPR) reaches 1\% and 0.1\%, respectively.
\begin{table}[h]\renewcommand{\arraystretch}{1.4}
\centering
\caption{Detection Rates with Different Disturbance Rates.}\label{Efficacy}
%\vspace{-0.1in}
\begin{tabular}{p{3cm}| p{2.1cm} p{2.1cm}}
% \diagbox{Height}{Group}{BMI}&\(<18\)&\(18\sim22\)&\(>22\)\\
 \toprule
 \ &Foursquare&Yelp  \\
 \hline
\ Disturbance Rate=0.1\%&30.8\%&31.7\%\\
\ Disturbance Rate=1.0\%&65.3\%&72.2\%\\
%\ \# of social ties&330,898&1,262,659\\
\bottomrule
\end{tabular}
\end{table}
%We also present the Precision-Recall curves in Figure \ref{PR-curve}.

%\begin{figure}
%\centering
%  \includegraphics[width=0.45 \textwidth]{Foursquare-joint-c30z20-1-PR.eps}
%  \caption{The Precision-Recall curve of identity theft detection via joint model on Foursquare dataset.}\label{F-j1-PR}
%  \includegraphics[width=0.45 \textwidth]{Yelp-joint-c30z20-1-PR.eps}
%  \caption{The Precision-Recall curve of identity theft detection via joint model on Yelp dataset.}\label{Y-j1-PR}
%\end{figure}

\vspace{-0.05in}
\subsection{Performance Comparison} \label{comparations}

\subsubsection{Representative Models}\label{subsub-mehodes-repreign}
We compare our joint model CBM to some representative models in OSNs.
In Table \ref{comparation-method}, we list the features of these models.
%\begin{table}[t]\renewcommand{\arraystretch}{1.2}
%\centering
%\caption{Characteristics of Different Models}\label{comparation-method}
%\vspace{-0.1in}
%\scalebox{1.0}{\begin{tabular}{p{1.9cm}|p{1.7cm}p{1.7cm}p{1.7cm}}
%\toprule
%&Online Behavior (UGC)&Offline Behavior (Check-in)&Composite Manner\\
%\hline
%\ CF-KDE&NO&YES&N/A\\
%\ LDA&YES&NO&N/A\\
%\ FUSED&YES&YES&FUSED\\
%\ JOINT &YES&YES&JOINT\\
%\bottomrule
%\end{tabular}
%}
%\vspace{-0.15in}
%\end{table}

\begin{table}[t]\renewcommand{\arraystretch}{1.2}
\centering
\caption{Behaviors Adopted in Different Models}\label{comparation-method}
\vspace{-0.1in}
\scalebox{0.99}{\begin{tabular}{p{1.1cm}|p{2.95cm}p{3.45cm}}
\toprule
&Online Behavior (UGC)&Offline Behavior (Check-in)\\
\hline
\ CF-KDE&NO&YES\\
\ LDA&YES&NO\\
\ FUSED&YES&YES\\
\ JOINT &YES&YES\\
\bottomrule
\end{tabular}
}
%\vspace{-0.1in}
\end{table}

\textbf{CF-KDE}.
Before presenting the CF-KDE model, we introduce the Mixture Kernel Density Estimate (MKDE) \cite{LichmanS14} to give a brief prior knowledge on Kernel Density Estimate (KDE).
MKDE is an individual-level spatial distribution model based on KDE.
It is a typical spatial model describing user's offline behavioral pattern.
In this model, it mainly utilizes a bivariate density function in the following equations to capture the spatial distribution for each user:
\begin{align}
& f_{KDE}\left (e|E,h \right )= \frac{1}{n}\sum\nolimits_{j=1}^{n}K_{h}\left ( e-e^{j} \right ),   \label{KDE} \\
& K_{h}\left ( x \right )=\frac{1}{2\pi h}\exp\left ( -\frac{1}{2}x^TH^{-1}x \right ), \mathbf{H}=\begin{pmatrix}
h & 0\\
0 & h
\end{pmatrix}, \label{KernelFunction}\\
& f_{MKDE}\left ( e|E,h \right ) = \alpha f_{KD}\left ( e|E_{1} \right )+\left ( 1-\alpha \right ) f_{KD}\left ( e|E_{2} \right ).\label{MKDE}
\end{align}
In Eq. (\ref{KDE}), $E=\{e^1,...,e^n\}$ is a set of historical behavioral records for a user and $e^j=<x,y>$ is a two-dimensional spatial location (i.e., offline behavior).
Eq. (\ref{KernelFunction}) is a kernel function and $\mathbf{H}$ is the bandwidth matrix.
MKDE adopts Eq. (\ref{MKDE}), where $E_1$ is a set of an individual's historical behavioral records (individual component), $E_2$ is a set of his/her friends' historical behavioral records (social component), and $\alpha$ is the weight variable for individual component.
To detect identity thieves, we compute a  surprise index $S_e$ in Eq. (\ref{Detect-MKDE}) for each behavior $e$, defined as the negative log-probability of individual $u$'s conducting  behavior $e$:
\begin{equation}
S_{e}=-\log f_{MKDE}\left ( e|E_{u},h_{u}\right ).
\label{Detect-MKDE}
\end{equation}
Furthermore, we can select the top-$N$ behaviors with the highest $S_e$ as suspicious behaviors.

%\emph{CF-KDE}.
%In this method, it makes an implement by combining use collaborative filtering (CF) and KDE model.
In MKDE model, it assumes that users tend to do like their friends in the same chance.
It has not quantified the potential influence of different friends.
Thus, we introduce a collaborative filtering method to improve the performance.
Based on the historical behavioral records, it establishes a user-venue matrix $\mathbf{R}_{|\mathbf{U}| \times |\mathbf{V}|}$, where $U$ and $V$ are the number of users and venues, respectively;
$\mathbf{R}_{ij} = 1$ if user $i$ has visited venue $j$ in the training set, otherwise $\mathbf{R}_{ij} = 0$.
We adopt a matrix factorization method with an objective function in Eq. (\ref{MF-target}) to obtain feature vectors for each user and venue:
\begin{equation}
L = \min_{\mathbf{U},\mathbf{V}} \frac{1}{2} \sum_{i=1}^{U} \sum_{j=1}^{V} (\mathbf{R}_{ij}-u_{i}^{T}v_{j})^{2}+\frac{\lambda_1}{2} \sum_{i=1}^{U} u_{i}^{T}u_{i}+\frac{\lambda_2}{2} \sum_{j=1}^{V} v_{j}^{T}v_{j}.\label{MF-target}
\end{equation}
Specifically, we let
\begin{center}
$u_{i}=(u_{i}^{(1)}, u_{i}^{(2)}, ..., u_{i}^{(k)})^T$ and $v_{j}=(v_{j}^{(1)}, v_{j}^{(2)}, ..., v_{j}^{(k)})^T$.
\end{center}
We adopt a stochastic gradient descent algorithm in Eqs.  (\ref{SGD1}) and (\ref{SGD2}) in the optimization process:
\begin{equation}
u_{i}^{(k)} \leftarrow u_{i}^{(k)} - \alpha (\sum\nolimits_{j=1}^{V} (\mathbf{R}_{ij}-u_{i}^{T}v_{j})v_{j}^{(k)}+\lambda_1 u_{i}^{(k)}),\label{SGD1}
\end{equation}
\begin{equation}
v_{j}^{(k)} \leftarrow v_{j}^{(k)} - \alpha (\sum\nolimits_{i=1}^{U} (\mathbf{R}_{ij}-u_{i}^{T}v_{j})u_{i}^{(k)}+\lambda_2 v_{j}^{(k)}).\label{SGD2}
\end{equation}
Consequently, we can figure out $\hat{\mathbf{R}}=\mathbf{U}^{T}\mathbf{V}$, and use $\hat{r}_{ij}=u_{i}^{T}v_{j}$ as the weight variable for the KDE model.
To detect anomalous behaviors, we use Eq. (\ref{CF-KDE}) to measure the surprising index for each behavior $e$:
\vspace{-0.05in}
\begin{equation}
S_{e}=-\log \frac{\sum_{j=1}^{n}\hat{r}_{uj}K_{h}\left ( e-e^{j} \right )}{\sum_{j=1}^{n}\hat{r}_{uj}}.   \label{CF-KDE}
\end{equation}
\vspace{-0.1in}

\noindent Furthermore, we can select the top-$N$ behaviors with the highest $S_e$ as suspicious behaviors.

\vspace{+0.05in}
\textbf{LDA}.
Latent Dirichlet Allocation (LDA) \cite{Blei2003LDA944919944937} is a classic topic model.
User's online behavior pattern can be denoted as the mixing proportions for topics.
We aggregate the UGC of each user and his/her friends in the training set as a document, then use LDA to obtain each user's historical topic distribution $\theta_{his}$.
To get their present behavioral topic distribution $\theta_{new}$ in the test set.
For each behavior, we count the number of words assigned to the $k$th topic, and denoted it as $n(k)$.
The $k$th component of the topic proportion vector can be computed in Eq. (\ref{present}):
\begin{equation}
\theta_{new}^{(k)}=\frac{n\left ( k \right )+\alpha}{\sum_{i=1}^{K}( n\left ( i \right )+\alpha ) },
\label{present}
\end{equation}
where $K$ is the number of topics, and $\alpha$ is a hyperparameter.
Specifically, we set $\alpha = 50/K$.

%To detect anomalous behaviors, we investigate the Jensen-Shannon (JS) divergence of a user's historical topic vector, and present it by Eqs.  (\ref{JS}), where $P=(p_{1},...,p_{K})$, $Q=(q_{1},...,q_{K})$ denote two distributions and $M=\frac{P+Q}{2}$.
%We can select the top-$N$ behaviors with the highest $D_{JS}(\theta_{his}, \theta_{new})$ as suspicious behaviors.
%\begin{equation}\label{KL}
%D_{KL}\left ( P,Q \right )=\sum\nolimits_{i=1}^{K}p_{i}\cdot \ln\frac{p_{i}}{q_{i}},
%\end{equation}
%\vspace{-0.15in}
%\begin{equation}\label{JS}
%D_{JS}\left ( P,Q \right )= \frac{1}{2}\left [ D_{KL} \left ( P, M \right )+ D_{KL} \left ( Q, M \right) \right].
%\end{equation}

To detect anomalous behaviors, we measure the distance between a user's historical and present topic distribution by using the Jensen-Shannon (JS) divergence in Eqs.  (\ref{KL}) and  (\ref{JS}):
\begin{equation}\label{KL}
D_{KL}\left ( \theta_{his},\theta_{new} \right )=\sum\nolimits_{i=1}^{K}\theta_{his}^{(i)}\cdot \ln \left({\theta_{his}^{(i)}}/{\theta_{new}^{(i)}}\right),
\end{equation}
\vspace{-0.1in}
\begin{equation}\label{JS}
D_{JS}\left ( \theta_{his},\theta_{new} \right )= \frac{1}{2}\left [ D_{KL} \left ( \theta_{his}, M \right )+ D_{KL} \left ( \theta_{new}, M \right) \right].
\end{equation}
where $M=\frac{\theta_{his}+\theta_{new}}{2}$.
We can select the top-$N$ behaviors with the highest $D_{JS}(\theta_{his}, \theta_{new})$ as suspicious behaviors.

\vspace{+0.05in}
\textbf{Fused Model}.
Egele et al. \cite{EgeleSKV17} propose COMPA which directly combining use users' explicit behavior features, e.g., language, links, message source, et al.
In our case, we setup a fused model, which deep combine users' implicit behavior features discovered by CF-KDE and LDA to detect identity theft.
We try different thresholds for the CF-KDE model and LDA model (i.e., different classifiers).
For each pair (i.e., a CF-KDE model and an LDA model), we treat any behavior that fails to pass either identification model as suspicious behavior, and compute true positive rate and false positive rate to draw the ROC curve and estimate the AUC value.

\vspace{-0.03in}
\subsubsection{Performance Comparison}
We compare the performance of our method with the typical ones in terms of \emph{detection efficacy} (AUC) and
\emph{response latency}.
The latter denotes the number of behaviors in the test set needed to cumulate for detecting a specific identity theft case.

\begin{figure}[t]
\centering
    \includegraphics[width=0.45 \textwidth]{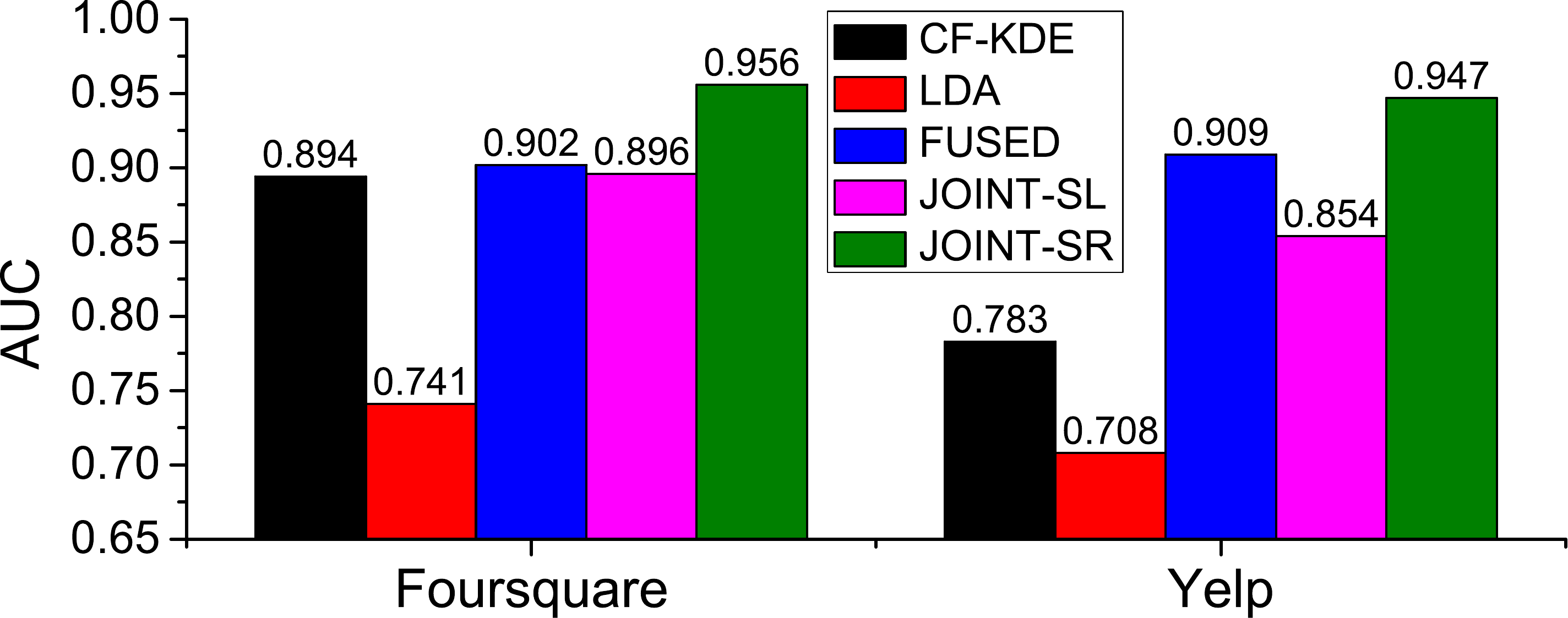}
% \vspace{-0.1in}
  \caption{Identity theft detection efficacy.}
  \label{Comparison-Result}
   %\vspace{-0.15in}
\end{figure}

\vspace{+0.02in}
\textbf{Detection Efficacy Analysis}.
In Fig. \ref{Comparison-Result}, we present the results of all comparison methods.
Our joint model outperforms all other methods on the two datasets.
The AUC value reaches $0.956$ and $0.947$ in Foursquare and Yelp datasets, respectively.

There are three reasons for its outstanding performance.
Firstly, it embraces different types of behaviors and exploits them in a unified model.
Secondly, it takes advantage of the community members' and friends' group-level behavior patterns to overcome the data insufficiency and concept drift \cite{BurszteinBMPAAPS14} in individual-level behavioral patterns.
Finally, it utilizes correlations among different behavioral spaces.

From the results, we have several other interesting observations:
(1) LDA model performs poor in both datasets which may indicate its performance is strongly sensitive to the data quality.
(2) CF-KDE and LDA model performs not well in Yelp dataset comparing to Foursquare dataset, but the fused model observes a surprising reversion.
(3) The joint model based on \emph{relative anomalous score} $S_r$ outperforms the model based on \emph{logarithmic anomalous score} $S_l$.
(4) The joint model (i.e., JOINT-SR, the joint model in the following sections all refer to the joint model based on $S_r$) is indeed superior to the fused model.

\begin{figure}[t]
\begin{center}
\begin{tabular}{cc}
%  \scalebox{0.5}{\includegraphics[width=0.45 \textwidth]{Compare-1-5-AUC-Foursquare.eps}}&
%  \scalebox{0.5}{\includegraphics[width=0.45 \textwidth]{Compare-1-5-AUC-Yelp.eps}}\\
  \scalebox{0.5}{\includegraphics[width=0.45 \textwidth]{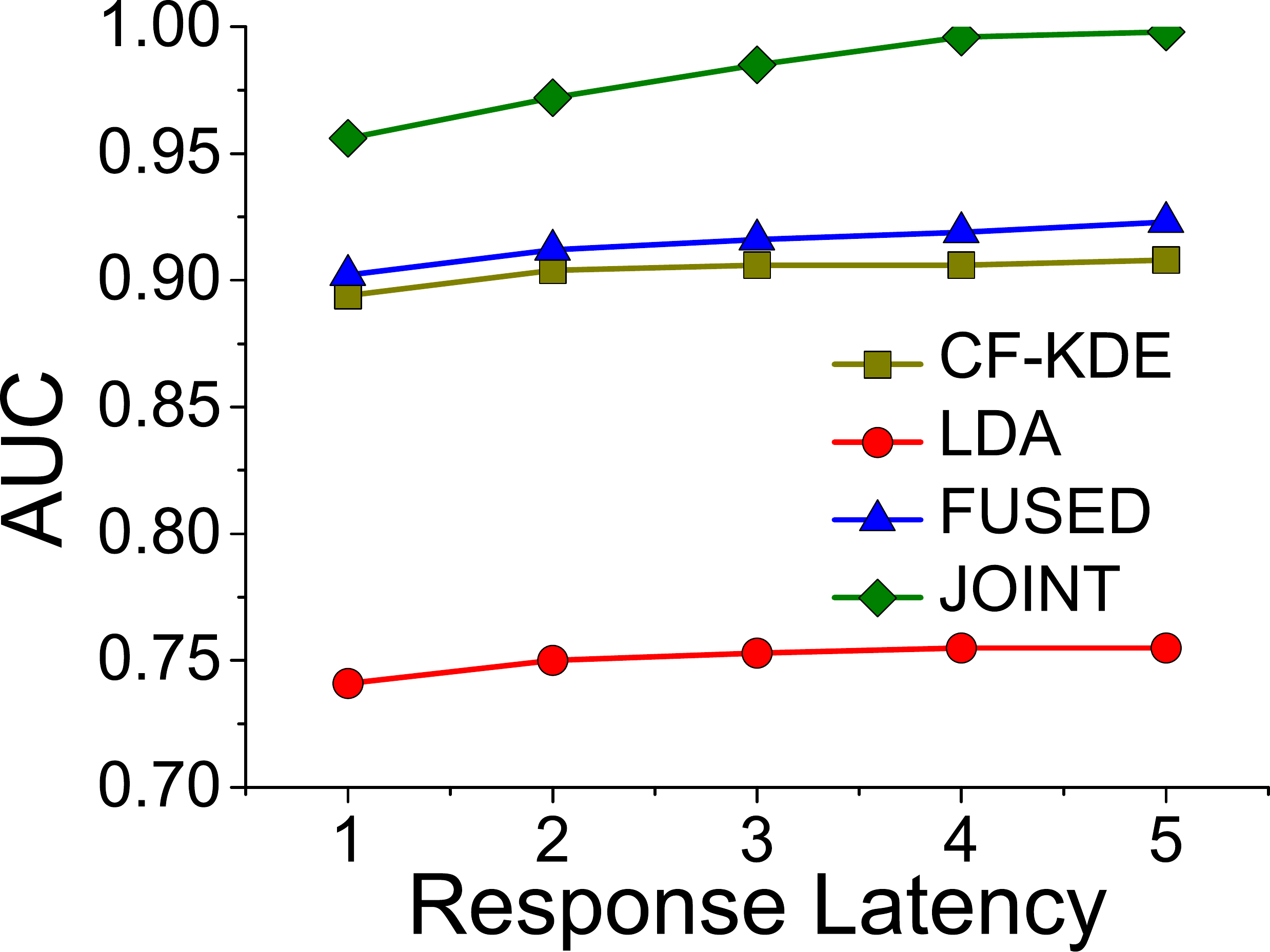}}&
  \scalebox{0.5}{\includegraphics[width=0.45 \textwidth]{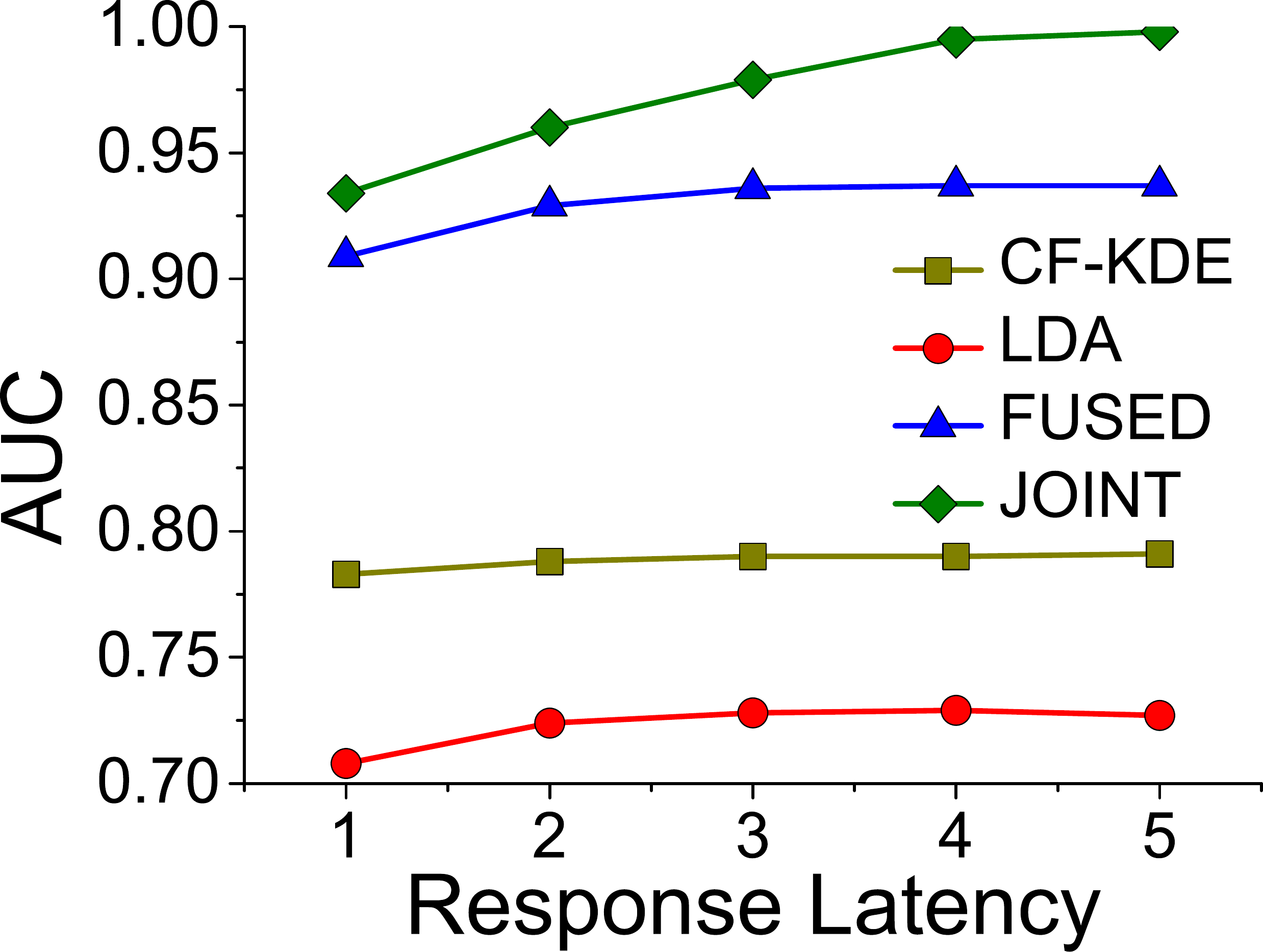}}\\
{\small(a) Foursquare} & {\small (b) Yelp}
\end{tabular}
\end{center}
  \vspace{-0.05in}
  \caption{Identity theft detection efficacy via different response latency (i.e., the number of behaviors in the test set we cumulated).}\label{Comparison-RT-Y}
 \vspace{-0.05in}
\end{figure}

\begin{figure}[t]
\begin{center}
\begin{tabular}{cc}
%  \scalebox{0.5}{\includegraphics[width=0.5 \textwidth]{Foursquare-TPR-1-5.eps}}&
%  \scalebox{0.5}{\includegraphics[width=0.5 \textwidth]{Yelp-TPR-1-5.eps}}\\
  \scalebox{0.5}{\includegraphics[width=0.45 \textwidth]{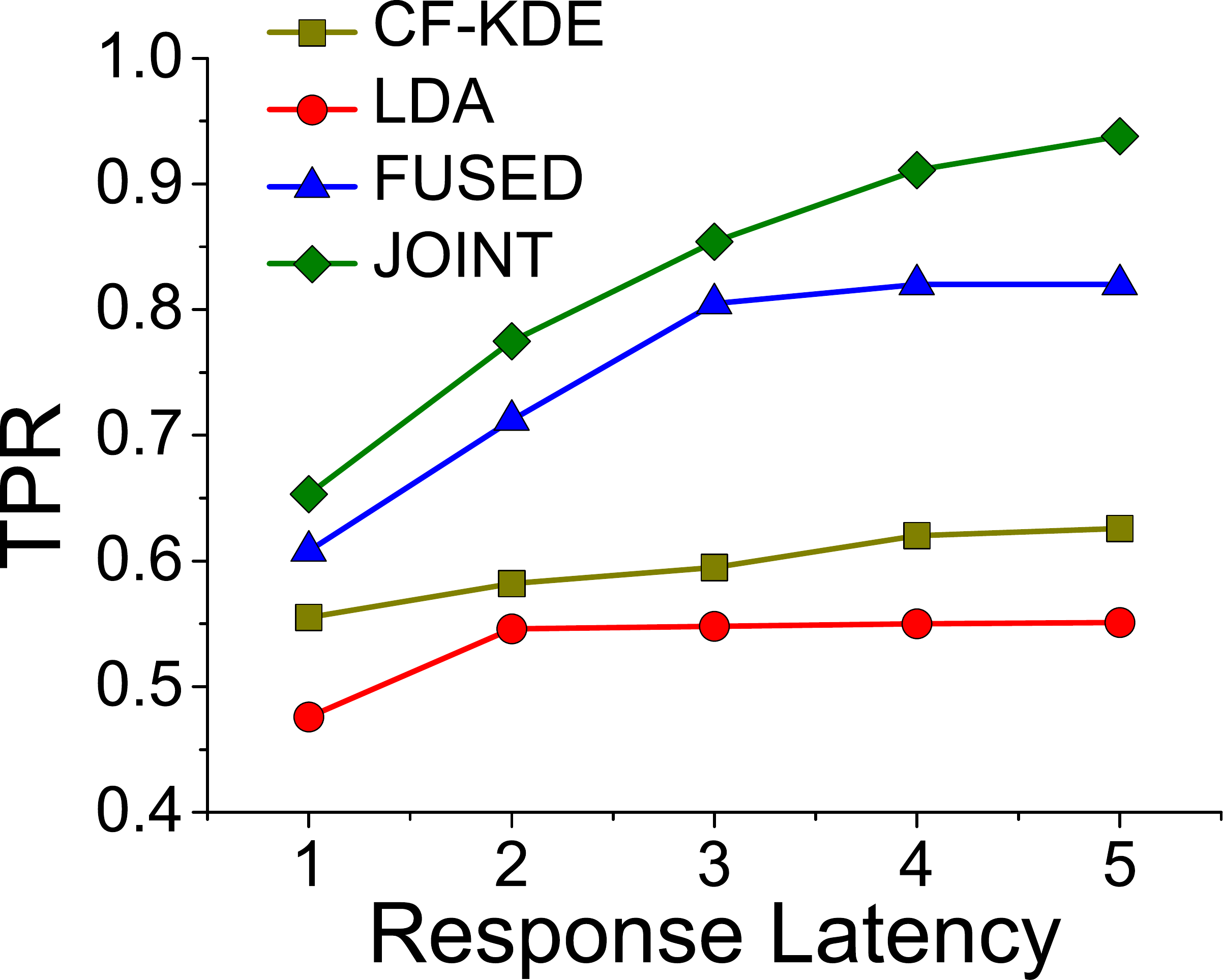}}&
  \scalebox{0.5}{\includegraphics[width=0.45 \textwidth]{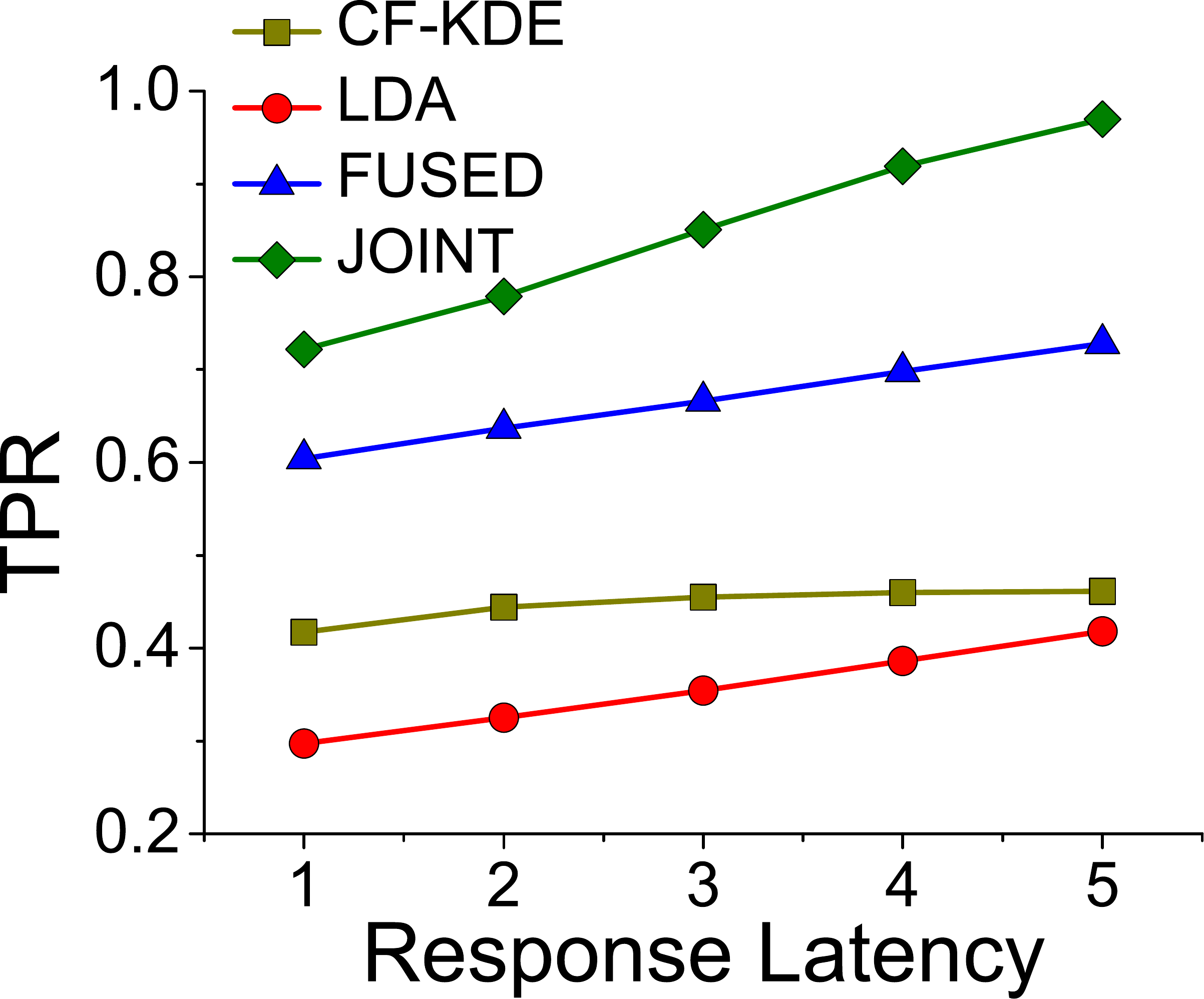}}\\
{\small(a) Foursquare} & {\small (b) Yelp}
\end{tabular}
\end{center}
 \vspace{-0.1in}
  \caption{The detection rates (TPR) via different response latency with disturbance rate=0.01 (FPR=0.01).}\label{Comparison-TPR-Y}
   \vspace{-0.1in}
\end{figure}

\vspace{+0.02in}
\textbf{Response Latency Analysis}.
For each model, we also evaluate the relationship between the efficacy and response latency (i.e., a response latency $k$ means that the identity theft is detected based on $k$ recent continuous behaviors).
Figs. \ref{Comparison-RT-Y} and \ref{Comparison-TPR-Y} demonstrate the AUC values and TPRs via different response latency in each model on both datasets.

The experimental results indicate that our joint model CBM is superior to all other methods.
The AUC values of our joint model can reach $0.998$ in both Foursquare and Yelp with $5$ test behavioral records.
The detection rates (TPR) of our joint model can reach $93.8\%$ in Foursquare and $97.0\%$ in Yelp with $5$ test behavioral records and disturbance rate (FPR) values $1.0\%$.

%\begin{figure}
%\centering
%    \includegraphics[width=0.45 \textwidth]{Compare-1-5-AUC-Yelp.eps}
%   % \includegraphics[width=0.45\textwidth]{imagesrvice-latency-180.eps}}
%  \caption{Identity Theft Detection Accuracy via different response time on Yelp.}
%  \label{Comparison-RT-Y}
%\end{figure}

\subsubsection{Robustness Analysis}
Generally,  there are two kinds of mutations in individual-level (IL) suspicious behavioral patterns:

\vspace{+0.02in}
\textbf{Completely Behavioral Mutation}.
Some thieves tore off their masks once intruding into victim's account.
They usually perform totally different interest in venues and topics.

\vspace{+0.02in}
\textbf{Partially Behavioral Mutation}.
Some extremely cunning thieves maintain part of victim's behavioral pattern to get further benefits from the victim's friends.
They may show partial behavior mutation which makes it harder to detect these anomalous behaviors.

\begin{figure}[t]
\centering
\includegraphics[width=0.48 \textwidth]{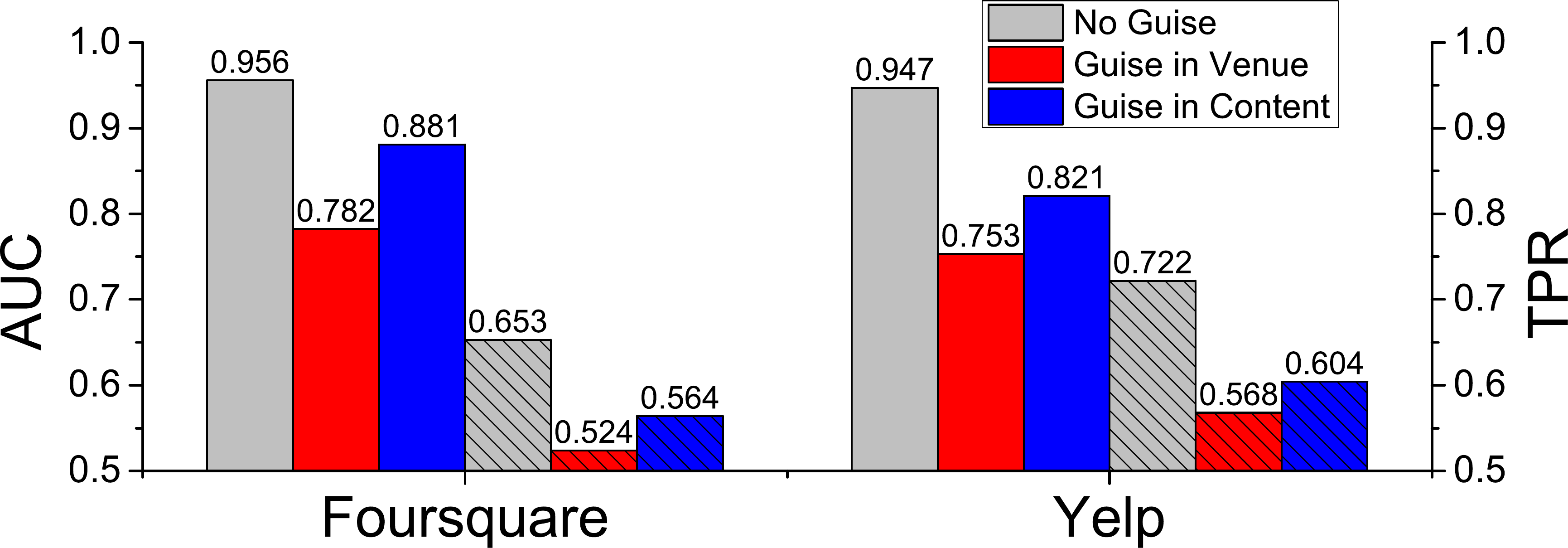}
% \vspace{-0.05in}
\caption{The efficacy (AUC) and detection rate (TPR) of identity theft detection via the joint model CBM in different scenarios with disturbance rate=0.01 (FPR=0.01). Painted ones denote AUC and shaded ones denote TPR.}\label{addition}
 \vspace{-0.1in}
\end{figure}

In the previous experiments, we evaluate the performance of our model in a scenario where thieves act like normal users by exchanging normal user's behavioral records (exchanging both venue and UGC).
Furthermore, we consider the harder scenarios where thieves know part of victim's habits and accordingly imitate victims.
%in the original pattern (i.e., partially behavioral mutation, which guise in venue or content)
%we only know a part of information in the test set (e.g., we only know the location or the UGC in the test set).
We apply our model to these scenarios, and demonstrate experimental results in Fig. \ref{addition}.
The results validate that our method is robust for coping with various suspicious behaviors.

%Since there are too many kinds of identity theft scenarios, we also simulate partial behavioral mutation as suspicious behaviors to validate the robustness of our model.
%We simulate two different kinds of partial behavioral mutations (venue mutation and UGC mutation).
%These mutations are harder to detect, however our model can still make good judgements.

%\begin{figure}
%\centering
%%\includegraphics[width=0.45 \textwidth]{addition-AUC.eps}
%\includegraphics[width=0.38 \textwidth]{addition-AUC.pdf}
%\caption{The efficacy of identity theft detection via joint model in different scenarios.}\label{AUC-addition}
%\end{figure}
%
%\begin{figure}
%%\includegraphics[width=0.45 \textwidth]{addition-TPR.eps}
%\centering
%\includegraphics[width=0.38 \textwidth]{addition-TPR.pdf}
%\caption{The detection rate (TPR) of identity theft detection via joint model in different scenarios with disturbance rate=0.01 (FPR=0.01).}\label{TPR-addition}
%\end{figure}

%\subsubsection{Comparison with Traditional Methods}\label{subsub-comparisnno-trationalmehods}

%\section{Discussion}\label{section-discussion}

\subsection{Explanations on Advantages of Joint Model}

\subsubsection{Intuitive Explanations}
%Fused model \cite{EgeleSKV17}, a composite behavior model, is a straightforward solution to detect identity theft.
%It first captures features in each behavior space respectively and then makes a fused metric based on these features in different dimensions.
%With the possible complementary effect among different behavior spaces, this fused model can act as a possibly feasible solution for the problem.
%However, the identification efficacy can be further improved, since the fused model neglects potential links among different spaces of behaviors.
Generally, there are two paradigms to integrate behavioral data: the \emph{fused} and \emph{joint} manners \cite{EgeleSKV17} .
Fused models \cite{EgeleSKV17} are a relatively simple and straightforward kind of composite behavior models.
They first capture  features in each behavior space respectively, and then make  a comprehensive metric based on these features in different dimensions.
With the possible complementary effect among different behavior spaces, they can act as a feasible solution for integration.
%%%%%%%%%%%%%%%%%%%%%%%%%%%%%%%%%%%%%%%%%%%%%%%%%%%%%%%%%%%%%%%%%%%%%%%%%%
However, the identification efficacy can be further improved, since  fused models neglect potential links among different spaces of behaviors.
We take an example where a person posted a picture in an OSN when he/she visited a park.
If this composite behavior is simply separated into two independent parts: he/she once posted a picture and he/she once visited a park, the difficulty in relocating him/her from a group of users is possibly increased, since there are more users satisfy these two simple conditions comparing to the original condition.

On the contrast, our joint model CBM sufficiently exploits the correlations between behaviors in different dimensions, then increases the certainty of users' behavior patterns, which contributes to a better identification efficacy.

\subsubsection{Theoretical Explanation}
We provide an underlying information theoretical explanation for the gain of joint models.
The well-known \emph{Chain Rule for Entropy} \cite{cover1991entropy},
%\begin{equation}\label{entropy}
\[H(X_{1}, X_{2}, ..., X_{N}) \leq \sum\nolimits_{i=1}^{N}H(X_i),\]
%\end{equation}
%The chain rule for entropy in Eq. (\ref{entropy})
indicates that the entropy of $N$ simultaneous events, denoted by $X_i$, $i=1,2, \cdots, N$, is no more than the sum of the entropies of each individual event, and are equal if the events are independent.

The chain rule for entropy  shows that the joint behavior has lower uncertainty comparing to the sum of the uncertainty in each component \cite{Song2010Limits}.
This can serve as a theoretical explanation of the advantages of our joint model.

\section{Literature Review}\label{section-Related}
%\section{Background and Related work}\label{section-Related}
%Background
%Identity theft is the deliberate use of other person's identity, usually as a method to gain a financial advantage or obtain credit and other benefits in other person's name.
%It may also result in other person's disadvantage or loss.
%According to the National Institute of Justice, identity theft became perhaps the defining crime of the information age, with an estimated nine million cases each year \cite{Newman2005Identity}.
%With the development of online social networks (OSNs), detecting identity theft cases in OSNs turned to be an ever-present and growing issue.
%Preventing and detecting identity theft are two important issues in OSN security.
%Related work

Recently, researchers found that users' behavior can identify their identities\cite{de2015unique,youyou2015computer,AbouelenienPMB17}.
Typically, behavior-based user identification include two phases: user profiling and user identifying:

\vspace{+0.02in}
\textbf{User profiling} is a process to characterize a user with his/her history behavioral data.
Some works focus on statistical characteristics to establish the user profile.
Naini et al. \cite{NainiUTV16} studied the task of identifying the users by matching the histograms of their data in the anonymous dataset with the histograms from the original dataset.
Egele et al. \cite{EgeleSKV17} proposed a behavior-based method to identify compromises of high-profile accounts.
Ruan et al. \cite{ProfilingOnlineSocialBehaviors} conducted a study on online user behavior by collecting and analyzing user clickstreams of a well known OSN.
Lesaege et al. \cite{LesaegeSLV16} developed a topic model extending the Latent Dirichlet Allocation (LDA) to identify the active users.
Viswanath et al. \cite{ViswanathBCGGKM14} presented a technique based on Principal Component Analysis (PCA) that accurately modeled the ``like'' behavior of normal users in Facebook and identified significant deviations from it as anomalous behaviors.
Tsikerdekis and Zeadally \cite{TsikerdekisZ14} presented a detection method based on nonverbal behavior for identity deception, which can be applied to many types of social media.
These methods above mainly concentrated on a specific dimension of the composite behavior without utilizing the correlations among multi-dimensional behavior data.

Vedran et al. \cite{sekara2016fundamental} explored the complex interaction between social and geospatial behavior and demonstrated that social behavior can be predicted with high precision.
Yin et al. \cite{YinHZWZHS16} proposed a probabilistic generative model combining use spatiotemporal data and semantic information to predict user's behavior.
These studies implied that composite behavior features are possibly helpful for user identification.

\vspace{+0.02in}
\textbf{User identifying} is a process to match the same user in two datasets or distinguish anomalous users/behaviors.
User identifying can be applied to a variety of tasks, such as detecting anomalous users or match users across different data sources.
Mazzawi et al. \cite{MazzawiDRENL17} presented a novel approach for detecting malicious user activity in databases by checking user's self-consistency and global-consistency.
Lee and Kim \cite{WarningBird} proposed a suspicious URL detection system for Twitter to detect users' anomalous behaviors.
%Thomas et al. \cite{URLSpamFiltering} leveraged Monarch's feature collection infrastructure to study distinctions among $11$ million URLs drawn from email and Twitter.
Cao et al. \cite{UncoveringMaliciousAccounts} designed and implemented a malicious account detection system for detecting both fake and compromised real user accounts.
Zhou et al. \cite{ZhouLZM16} proposed an FRUI algorithm to match users among multiple OSNs.
These works mainly detected the population-level anomalous behaviors which indicated strongly difference from other behaviors.
While, they did not consider that the individual-level coherence of users' behavioral patterns can be utilized to detect online identity thieves.

\vspace{+0.03in}
%\vspace{-0.05in}
\section{Conclusion and Future Work}\label{section-Conclusion}
\vspace{+0.02in}
We investigate the feasibility in building a ladder from low-quality behavioral data to a high-quality behavioral model for user identification in online social networks (OSNs).
By exploiting the complementary effect among OSN users' multi-dimensional behaviors, we propose a joint probabilistic generative model  by integrating online and offline behaviors.
When the designed model is applied into identity theft detection in OSNs, its comprehensive performance, in terms of  the detection efficacy, response latency and robustness, is validated by  extensive evaluations implemented on real-life OSN datasets.
%This work would give the cybersecurity research community new insights into whether and how
%a real-time online identity authentication can be improved via modeling users' composite behavioral patterns.
This study gives new insights into whether and  how modeling users' composite behavioral patterns can improve online identity authentication.

Our behavior-based module mainly aims at detecting identity thieves after the the access control of account is broken.
It is not exclusive to the traditional methods for preventing identity theft.
On the contrary, it is easy to incorporate our module into traditional methods to solve identity theft problem better, since our method is non-intrusive and continuous.
We would like to leave the study on  the combinations of preventing and detecting methods from a systematic perspective  as future work.

\end{document}